\documentclass[prd,amsmath,amssymb,showpacs,superscriptaddress,nofootinbib,twocolumn]{revtex4-1}
\usepackage{multirow}
\usepackage{float}
\usepackage{graphicx}
\usepackage{dcolumn}
\usepackage{bm}
\usepackage{rotating}
\usepackage{color}
\usepackage{subfigure}
\usepackage{pstricks}
\usepackage{soul}
\usepackage{overpic}
\usepackage[english]{babel}
\usepackage[colorlinks,urlcolor=black,linkcolor=blue,anchorcolor=blue,citecolor=blue]{hyperref}
\usepackage{tabularx}
\usepackage{lineno}
\newcommand{\PreserveBackslash}[1]{\let\temp=\\#1\let\\=\temp}
\newcolumntype{C}[1]{>{\PreserveBackslash\centering}p{#1}}
\newcolumntype{R}[1]{>{\PreserveBackslash\raggedleft}p{#1}}
\newcolumntype{L}[1]{>{\PreserveBackslash\raggedright}p{#1}}

\newcommand{\BR}{{\cal B}}
\newcommand{\ee}{e^{+}e^{-}}
\newcommand{\mumu}{\mu^+\mu^-}

\newcommand{\jpsi}{J/\psi}

\newcommand{\kshort}{K^0_S}

\begin{document}

\title{\boldmath First measurement of $\ee \to pK^{0}_{S}\bar{n}K^{-} + c.c.$ above open charm threshold}

\author{
\begin{small}
\begin{center}
M.~Ablikim$^{1}$, M.~N.~Achasov$^{10,d}$, S. ~Ahmed$^{15}$, M.~Albrecht$^{4}$, M.~Alekseev$^{55A,55C}$, A.~Amoroso$^{55A,55C}$, F.~F.~An$^{1}$, Q.~An$^{52,42}$, J.~Z.~Bai$^{1}$, Y.~Bai$^{41}$
% ...
\\
\vspace{0.2cm}
(BESIII Collaboration)\\
\vspace{0.2cm} {\it
$^{1}$ Institute of High Energy Physics, Beijing 100049, People's Republic of China\\
% ...
$^{4}$ Bochum Ruhr-University, D-44780 Bochum, Germany\\
% ...
$^{10}$ G.I. Budker Institute of Nuclear Physics SB RAS (BINP), Novosibirsk 630090, Russia\\
% ...
$^{15}$ Helmholtz Institute Mainz, Johann-Joachim-Becher-Weg 45, D-55099 Mainz, Germany\\
% ...
$^{41}$ Southeast University, Nanjing 211100, People's Republic of China\\
$^{42}$ State Key Laboratory of Particle Detection and Electronics, Beijing 100049, Hefei 230026, People's Republic of China\\
% ...
$^{52}$ University of Science and Technology of China, Hefei 230026, People's Republic of China\\
% ...
$^{55}$ (A)University of Turin, I-10125, Turin, Italy; (B)University of Eastern Piedmont, I-15121, Alessandria, Italy; (C)INFN, I-10125, Turin, Italy\\
% ...
$^{d}$ Also at the Novosibirsk State University, Novosibirsk, 630090, Russia\\
% ...
\vspace{0.4cm}
}\end{center}
\end{small}
}
\noaffiliation

\date{\today}

\begin{abstract}
  The process $e^+e^-\rightarrow pK^{0}_{S}\bar{n}K^{-} + c.c.$ and
  its intermediate processes are studied for the first time, using data
  samples collected with the BESIII detector at BEPCII
  at center-of-mass energies of 3.773, 4.008,
  4.226, 4.258, 4.358, 4.416, and 4.600 GeV, with a total
  integrated luminosity of 7.4 fb$^{-1}$.  The Born cross section of
  $\ee \to p\kshort\bar{n}K^- + c.c.$ is measured at each
  center-of-mass energy, but no significant resonant structure in the
  measured cross-section line shape between 3.773 and 4.600 GeV is observed. No evident
  structure is detected in the $pK^-$, $nK^{0}_S$, $pK^0_{S}$, $nK^+$,
  $p\bar{n}$, or $\kshort K^-$ invariant mass distributions except for
  $\Lambda(1520)$.  The Born cross sections of
  $e^+e^-\rightarrow\Lambda(1520)\bar{n}K^{0}_{S} + c.c.$ and
  $e^+e^-\rightarrow \Lambda(1520)\bar{p}K^{+} + c.c.$ are
  measured, and the 90\% confidence level upper limits on the Born
  cross sections of $e^+e^-\rightarrow\Lambda(1520)\bar{\Lambda}(1520)$ are
  determined at the seven center-of-mass energies.
  There is an evident difference in line shape and magnitude of
  the measured cross sections between $\ee\to\Lambda(1520)(\to pK^-)\bar{n}K^{0}_{S}$
  and $\ee\to pK^-\bar{\Lambda}(1520)(\to\bar{n}K^{0}_{S})$.

\end{abstract}

\maketitle

\section{\boldmath Introduction}

The experimental discovery of unexpected resonances has brought new
opportunities to the study of quantum chromodynamics in the
charmonium and bottomonium energy regions~\cite{2008ss, 2011nb, 2014nb}. The state $Y(4260)$ was
discovered by the BaBar collaboration~\cite{2005rm, Lees:2012cn} in the initial
state radiation (ISR) process $\ee\to\gamma_{\rm ISR}\pi^+\pi^-\jpsi$
and confirmed by the CLEO~\cite{2006kg} and Belle~\cite{2007sj} collaborations in the
same process. This state was further confirmed by BESIII~\cite{ref6}
and again by Belle~\cite{ref7}.
Located above the $D\bar{D}$ mass
threshold, the $Y(4260)$ with $J^{PC}=1^{--}$ anomalously couples to
the hidden-charm final state $\pi\pi\jpsi$~\cite{2006ss}.
The same phenomenon had been observed in other $Y$
states, such as the $Y(4360)$ and $Y(4660)$~\cite{2008ss}.
Just recently, BESIII first reported that two structures around 4.22 GeV and 4.39 GeV
in $\ee$ line shape strongly couple to $\pi^+D^0D^{*-}$~\cite{ddpi-bes}.
These interesting but little-known phenomenons have prompted researchers to focus on this charmonium-like
spectroscopy~\cite{2008ss, 2011nb, 2014nb}.

Light hadron decays of $Y$ states have not been found above 4 GeV, nor have any such decays of charmonium resonances.
The continued search for light hadron decays helps further the understanding of the nature of
undefined states and charmonium resonances.

In this paper, we report the cross sections of $\ee\to
p\kshort\bar{n}K^- + c.c.$ and search for possible structures, such as
$Y$ states or higher charmonia , in the $\ee\to
p\kshort\bar{n}K^- + c.c.$ cross section line shape, using data
samples with a total integrated luminosity of 7.4 fb$^{-1}$ collected
with the BESIII detector at center-of-mass (c.m.) energies between
3.773 and 4.6 GeV.  All possible intermediate states in the
$pK^{-}$, $nK_{S}^{0}$, $pK_{S}^{0}$, $nK^{+}$, $p\bar{n}$, and $\kshort
K^-$ invariant mass spectra and charge conjugated modes, such as
$\Lambda^{*}$, $\Sigma^{*}$, $a_{0}(980)^+$, and other excited or
exotic states, are searched for in the
$e^{+}e^{-}\rightarrow~pK_{S}^0\bar{n}K^{-} + c.c.$ process.  However,
no significant structures, except for $\Lambda(1520)$, are seen in
any of the studied mass spectra.  The Born cross sections of
$e^+e^-\rightarrow\Lambda(1520)\bar{n}K^{0}_{S}$ and
$\Lambda(1520)\bar{p}K^{+} + c.c.$ are measured.  In the following
analysis, charged conjugated modes are included unless otherwise
indicated.  The process $e^+e^-\rightarrow~pK^{0}_{S}\bar{n}K^{-}$
with all of the potential intermediate states included is denoted as
the ``$p\kshort\bar{n}K^-$ mode" hereinafter.  Similarly, the
processes $\ee\rightarrow\Lambda(1520)\bar{n}K^{0}_{S}$,
$\Lambda(1520)\bar{p}K^{+}$ and $\Lambda(1520)\bar\Lambda(1520)$
are denoted as ``$\Lambda(1520)$ modes".

\section{\boldmath Experimental data and monte carlo Simulation}

The Beijing Electron Positron Collider II (BEPCII)~\cite{BESIII},
a double-ring electron--positron collider with a peak luminosity of
$10^{33}~{\rm cm}^{-2}{\rm s}^{-1}$ at the c.m.\ energy of 3.770 GeV
at a beam current of 0.93~A, operates  in the center-of-mass energy region from 2 to 4.6 GeV.
The Beijing Spectrometer III (BESIII)~\cite{BESIII}, which operates at the BEPCII storage ring,
covers 93$\%$ of 4$\pi$ solid angle. A small-cell
helium-gas-based main drift chamber (MDC) provides charged particle
momentum and ionization loss ($dE/dx$) measurements for charged
particle identification (PID).  The momentum resolution is better than
0.5\% at 1 GeV/$c$ in a magnetic field of 1~T.  A plastic scintillator
time-of-flight (TOF) system with a time resolution of 80 ps (110 ps)
for the barrel (end-caps) is utilized for additional charged particle
identification. A CsI(Tl) crystal electromagnetic calorimeter obtains a photon energy resolution of
2.5\% (5\%) at 1 GeV in the barrel (end-caps). A resistive-plate-counter muon system provides a
position resolution of 2 cm and can detect muon tracks with momenta
greater than 0.5 GeV/$c$.

All data samples used in this analysis are listed in
Table~\ref{tab:neu}.  The c.m.\ energies are measured using the
dimuon process $\ee\to(\gamma_{\rm ISR/FSR})\mumu$ with an
uncertainty of $0.8$ MeV~\cite{cmsenergy}, and the integrated
luminosities are measured with large-angle Bhabha scattering events with an
uncertainty of $1.0\%$~\cite{lum3770,lum3770new,luminosity}, where FSR denotes
final state radiation.  The data sets with c.m.\ energy above 4
GeV are named ``$XYZ$ data".

The {\sc geant4}-based~\cite{geant4} Monte Carlo (MC) simulation
framework {\sc boost}~\cite{boost}, consisting of event generators,
the detector geometry,  and detector
response, is used to evaluate the detector efficiency, estimate ISR
correction, optimize event selection criteria, and analyze
backgrounds. The effects of FSR are simulated by the {\sc photos}~\cite{photos} package.
Exclusive phase space (PHSP) MC samples for signal modes are modeled
with {\sc kkmc}~\cite{kkmc:pingrg,kkmc1,kkmc2} and {\sc
besevtgen}~\cite{besevtgen,evtgen}, which consider ISR effects and beam energy spreads.
Seven `inclusive' MC simulated data samples at the different c.m.\ energies, equivalent to the
respective integrated luminosity of each data set, are produced to
investigate potential backgrounds.  The main known processes and decay modes are
generated by {\sc besevtgen} with cross sections or branching
fractions obtained from the Particle Data Group (PDG)~\cite{PDG}, and
the remaining unmeasured events associated with charmonium decays or
open charm processes are generated with {\sc
  lundcharm}~\cite{besevtgen,Chen:2000tv}, while continuum light hadronic events
are generated with {\sc pythia}~\cite{pythia}.

\section{\boldmath Data analysis}

In this analysis, the $n$ and $\bar{n}$ candidates are not reconstructed,
and a partial reconstruction technique is employed to select the signal events of interest,
\textit{i.e.}, we reconstruct $p,~K^{0}_{S}(\to\pi^+\pi^-)$, and
$K^{-}$ only, while $\bar{n}$ is treated as a missing particle.
The presence of a $\bar{n}$ is inferred using the mass recoiling against the $pK^0_S K^-$ system, $M^{\rm rec}(pK^0_SK^-)=\sqrt{(E-\sum{E_i})^2-(\sum\overrightarrow{P_{i}})^2 }$ $(i = p, K^{0}_{S},K^-)$,
where $E$ is the c.m.\ energy, $E_i$ is the energy of the $i$th track, and $\sum\overrightarrow{P_{i}}$ is the vector
sum of the track momenta.
For signal candidate events, the distribution of $M^{\rm rec}$ peaks at the $\bar{n}$  nominal mass~\cite{PDG}.

Events with at least two positively charged tracks and two negatively charged
tracks are selected. All charged tracks must be well
reconstructed in the MDC with $|{\rm cos} \theta|~<~0.93$, where
$\theta$ is the polar angle between the charged track and the
positron beam direction.

The $\kshort$ candidate is reconstructed with a pair of oppositely charged pions,
where the point of the closest approach to the $\ee$ interaction point
is required to be within $\pm20$ cm in the beam direction.
A charged track is identified as a pion by using the
combined TOF and $dE/dx$ information.
To suppress random combinatorial backgrounds,
we require that the $\pi^+ \pi^-$ pair satisfies a secondary vertex fit~\cite{sec-vtx}, and the decay length,
which is the distance between production and decay vertexes,
is required to be greater than twice its resolution.  If there is more than one
$\pi^+\pi^-$ combination in an event, the one with the smallest $\chi^2$ of the secondary vertex
fit is retained.
A $\kshort$ signal is required to have the
$\pi^+\pi^-$ invariant mass within $|M_{\pi^+\pi^-}-m_{\kshort}|\le10$ MeV/c$^{2}$, where
$m_{\kshort}$ is the $\kshort$ nominal mass~\cite{PDG}.

After the selection of the two $\pi$ tracks from $\kshort$ decays, the
remaining charged tracks are assumed to be $p$ or $K$, and the point
of closest approach of these tracks to the $\ee$ interaction point
must be within $\pm10$~cm in the beam direction and within 1~cm in the
plane perpendicular to the beam.  The net charge of the $p$ and $K$
combination must be zero, and multiple combinations of $p$ and $K$ are
permitted.

\subsection{\boldmath $p\kshort\bar{n}K^-$ mode}

\begin{figure*}[htbp]
    \centering
    \mbox{
    \hskip -0.58cm
    \begin{overpic}[width=5.2cm,height=4.0cm,angle=0]{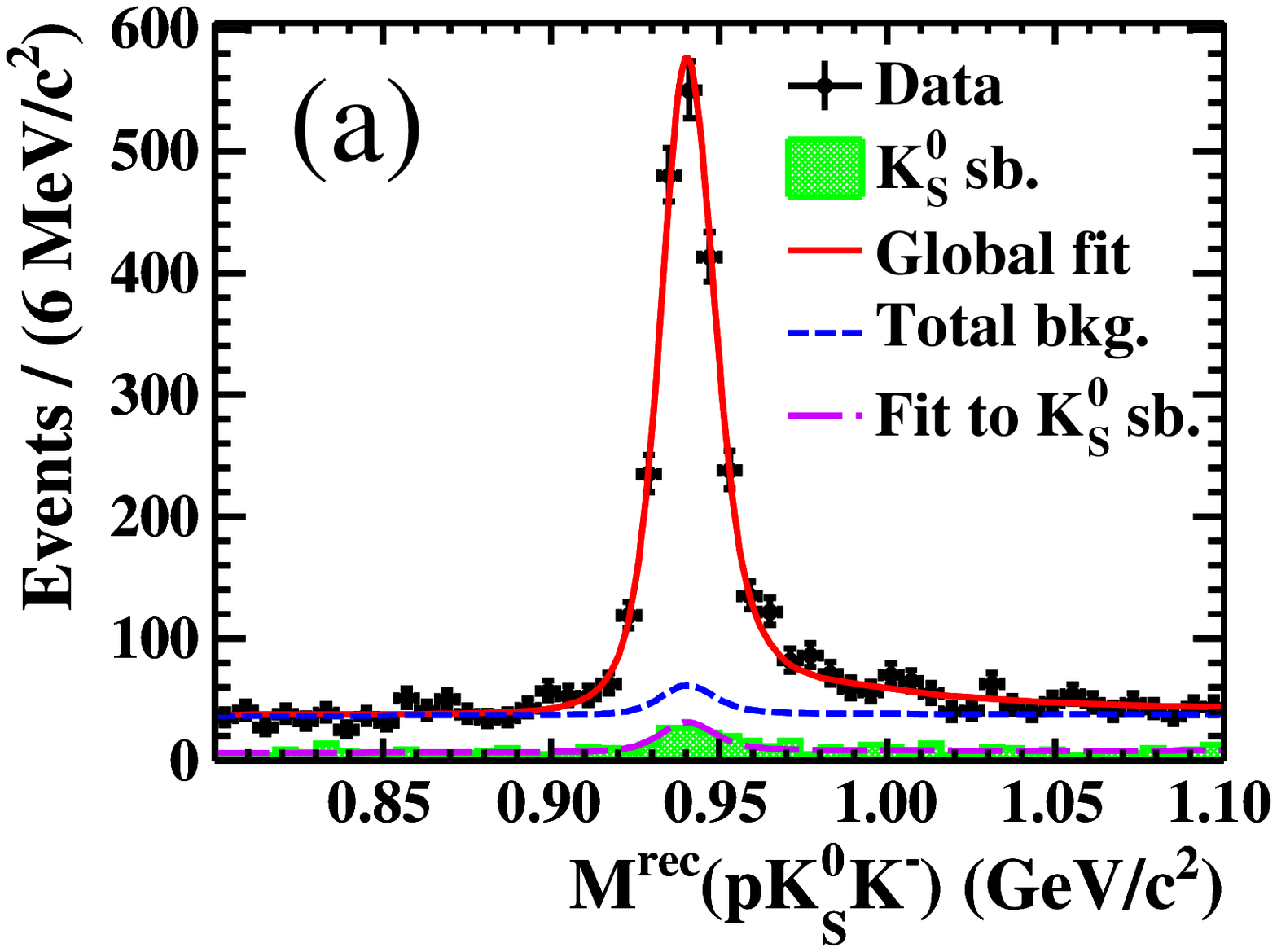}
    \end{overpic}
    \hskip -0.67cm
    \begin{overpic}[width=5.2cm,height=4.0cm,angle=0]{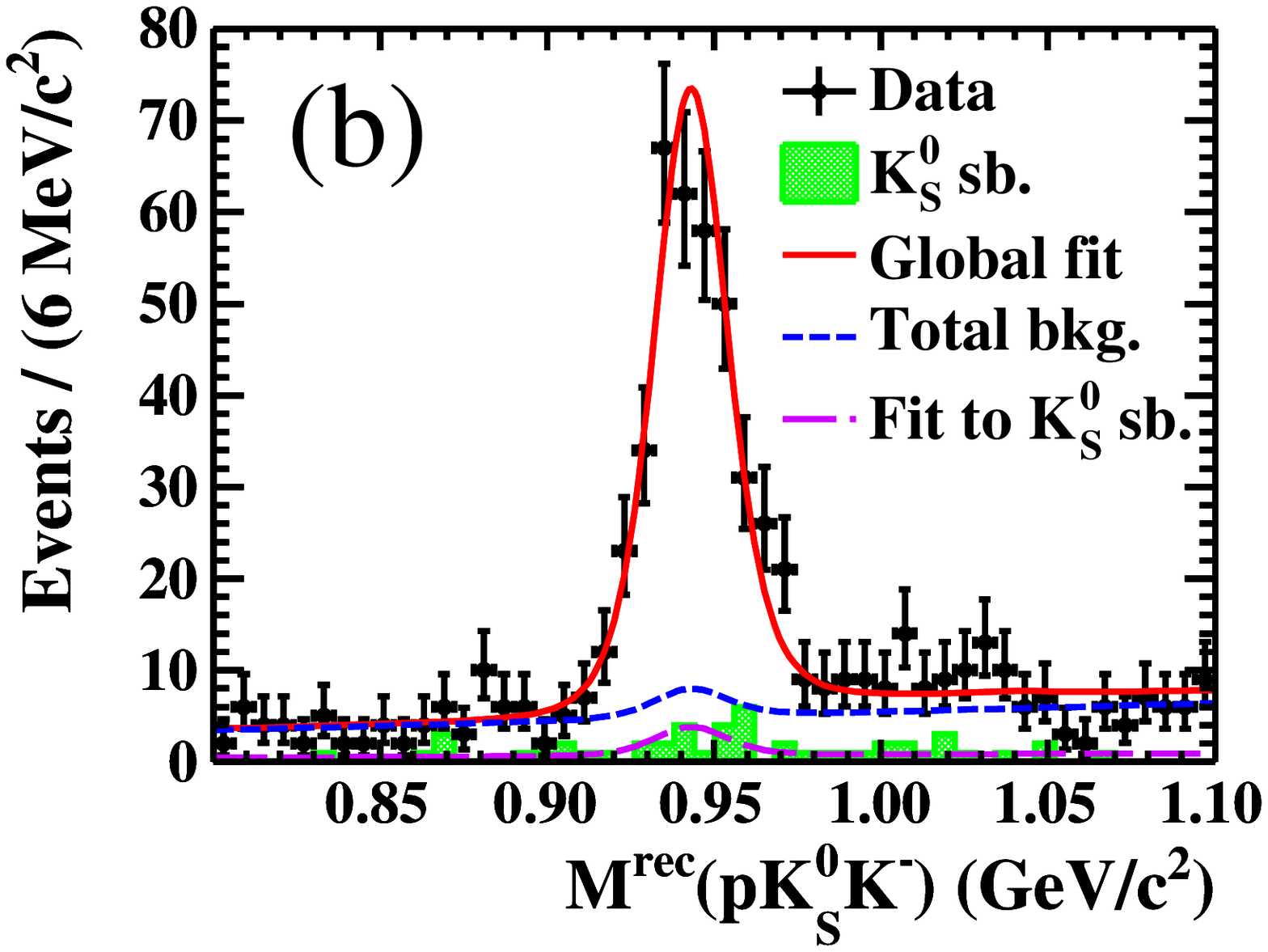}
    \end{overpic}
    \hskip -0.67cm
    \begin{overpic}[width=5.2cm,height=4.0cm,angle=0]{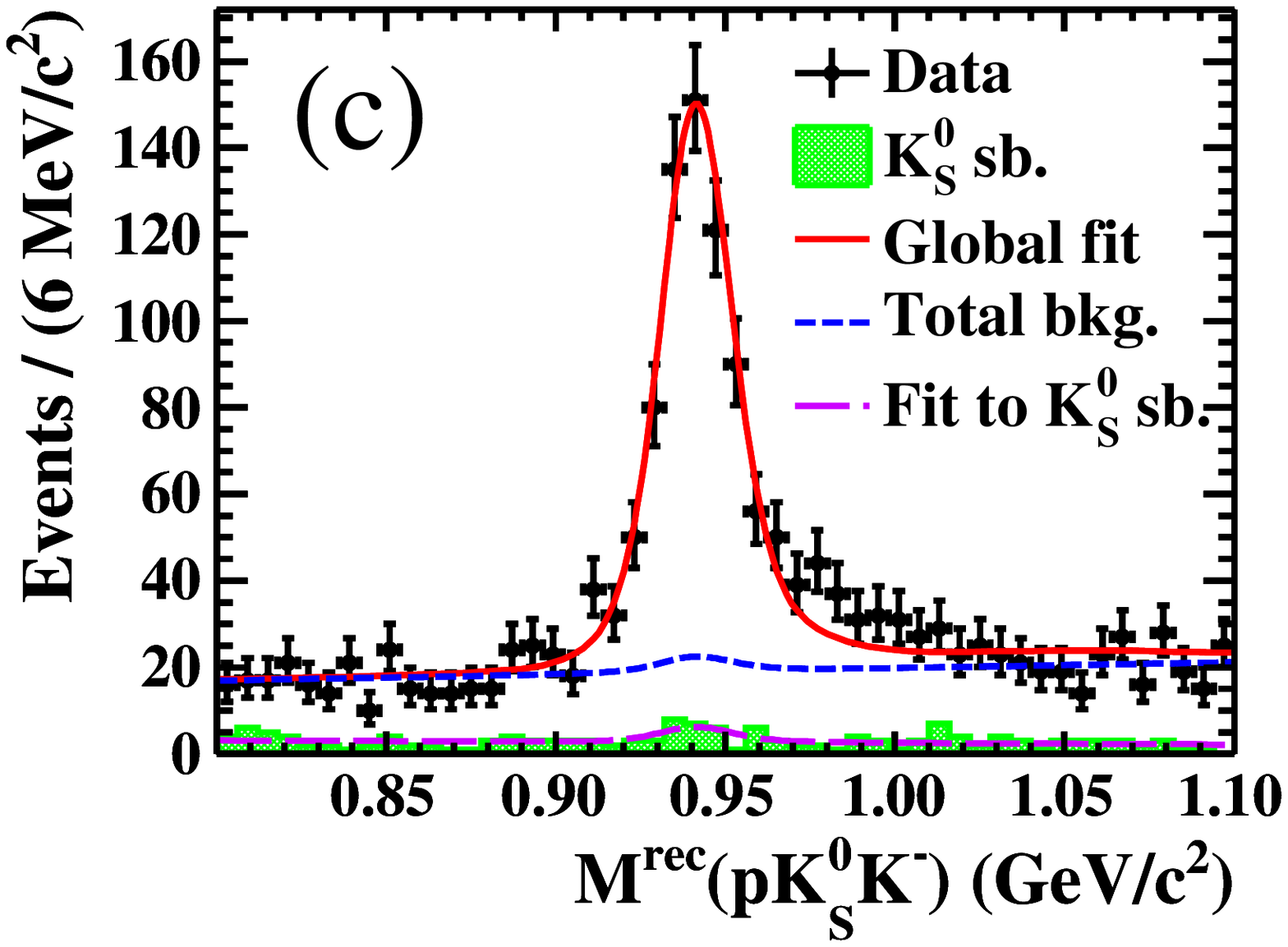}
    \end{overpic}
    \hskip -0.67cm
    \begin{overpic}[width=5.2cm,height=4.0cm,angle=0]{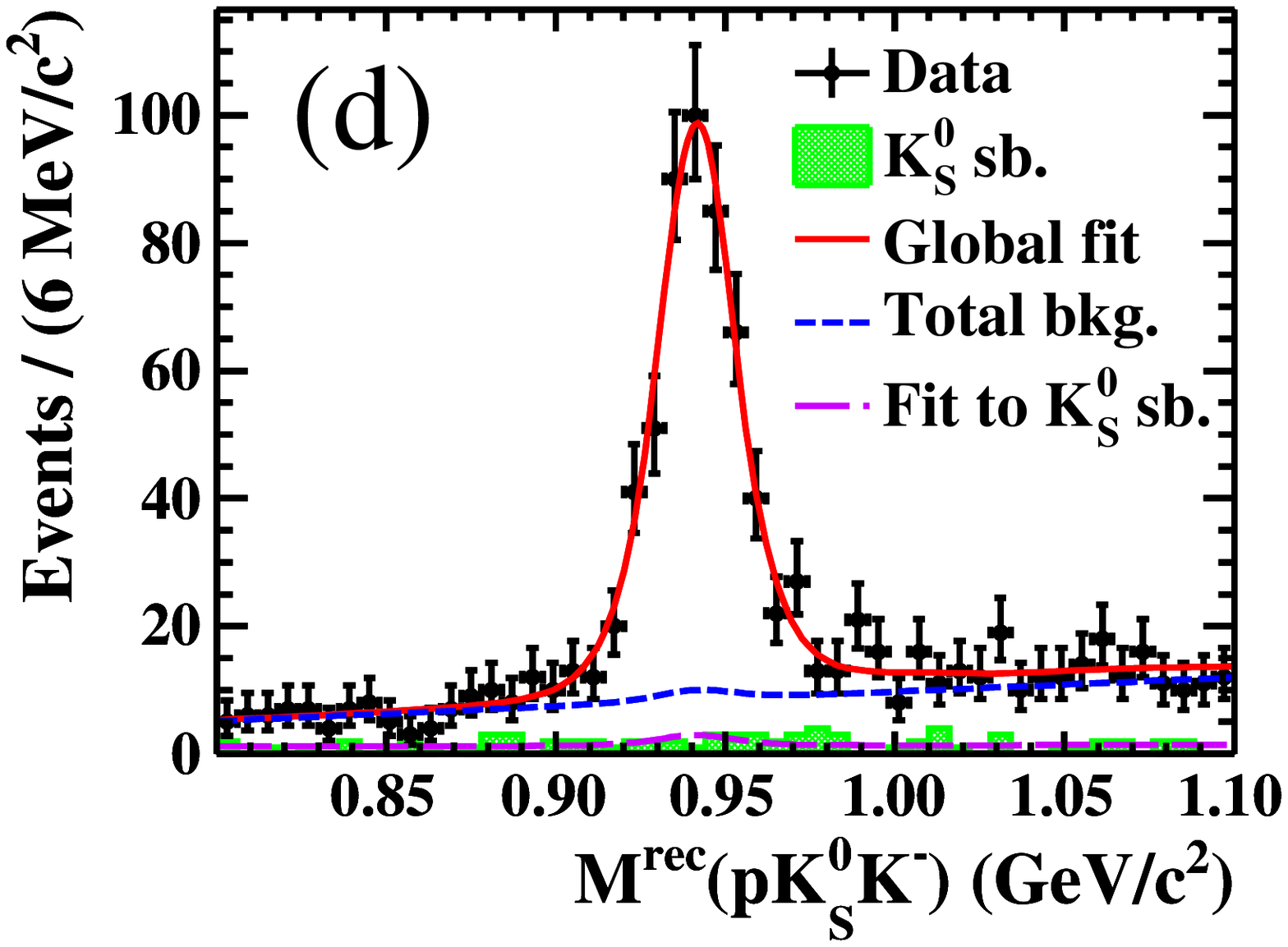}
    \end{overpic}
    }
    \hskip -0.4cm \mbox{
    \begin{overpic}[width=5.2cm,height=4.0cm,angle=0]{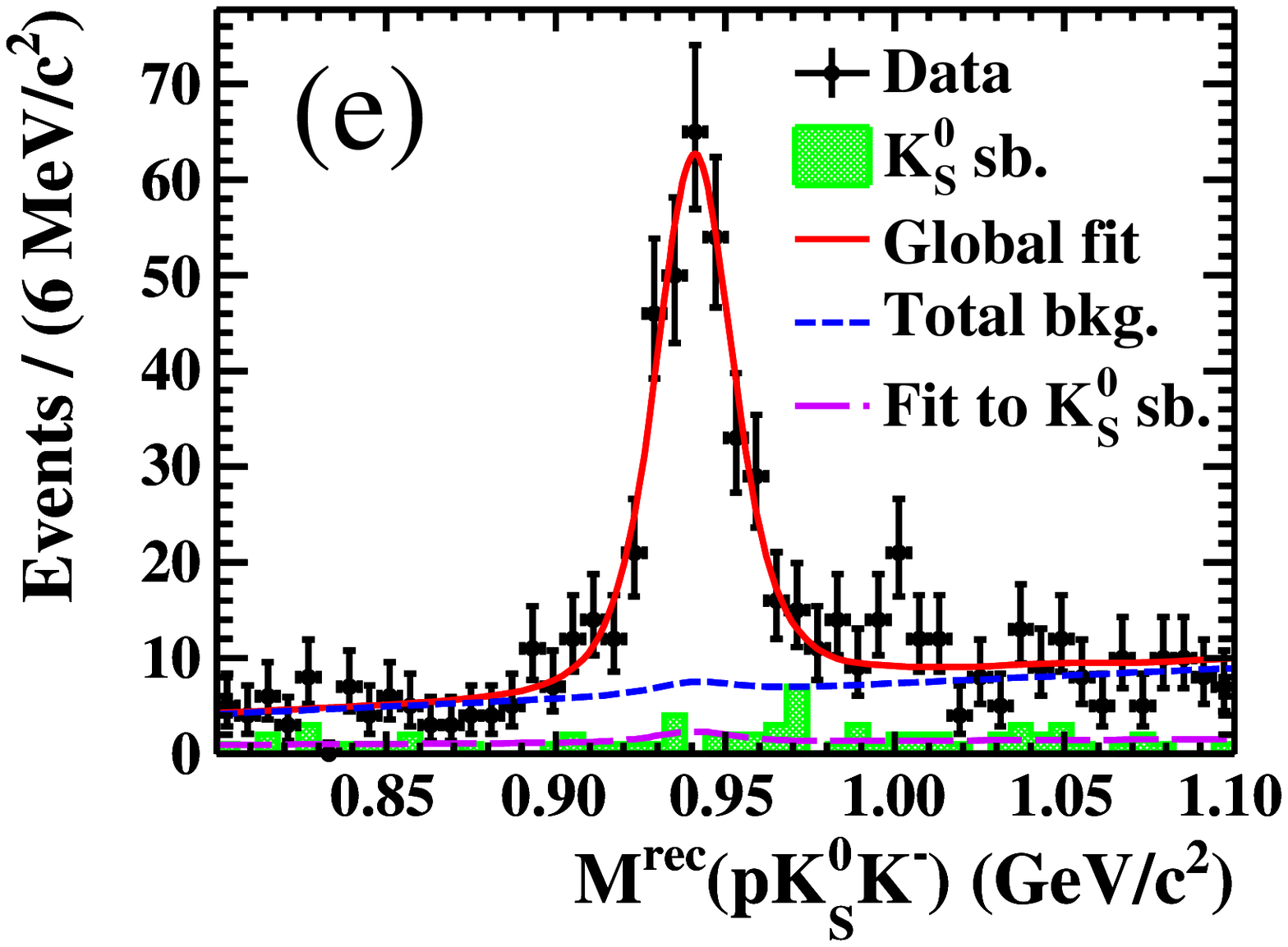}
    \end{overpic}
    \begin{overpic}[width=5.2cm,height=4.0cm,angle=0]{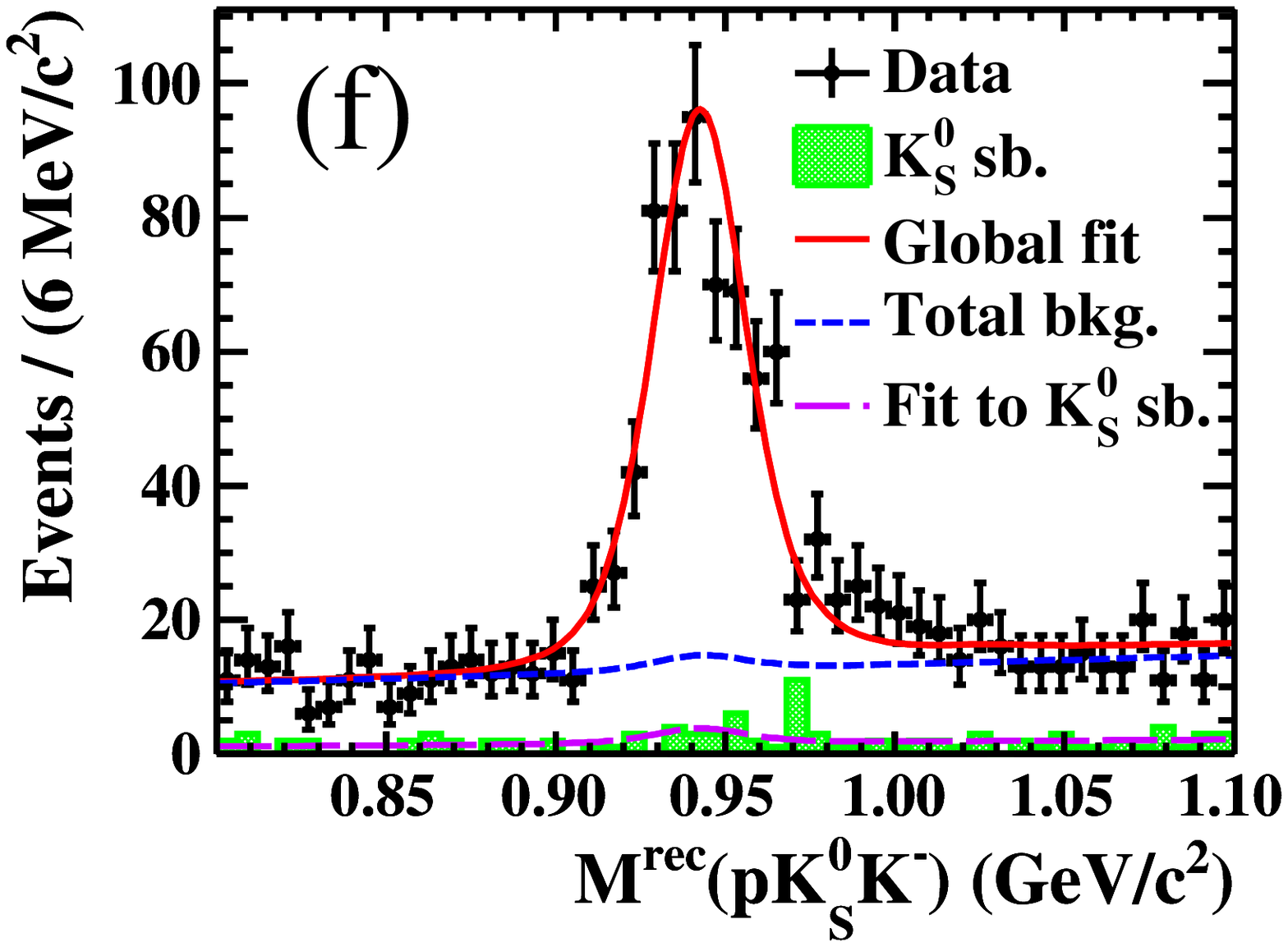}
    \end{overpic}
    \begin{overpic}[width=5.2cm,height=4.0cm,angle=0]{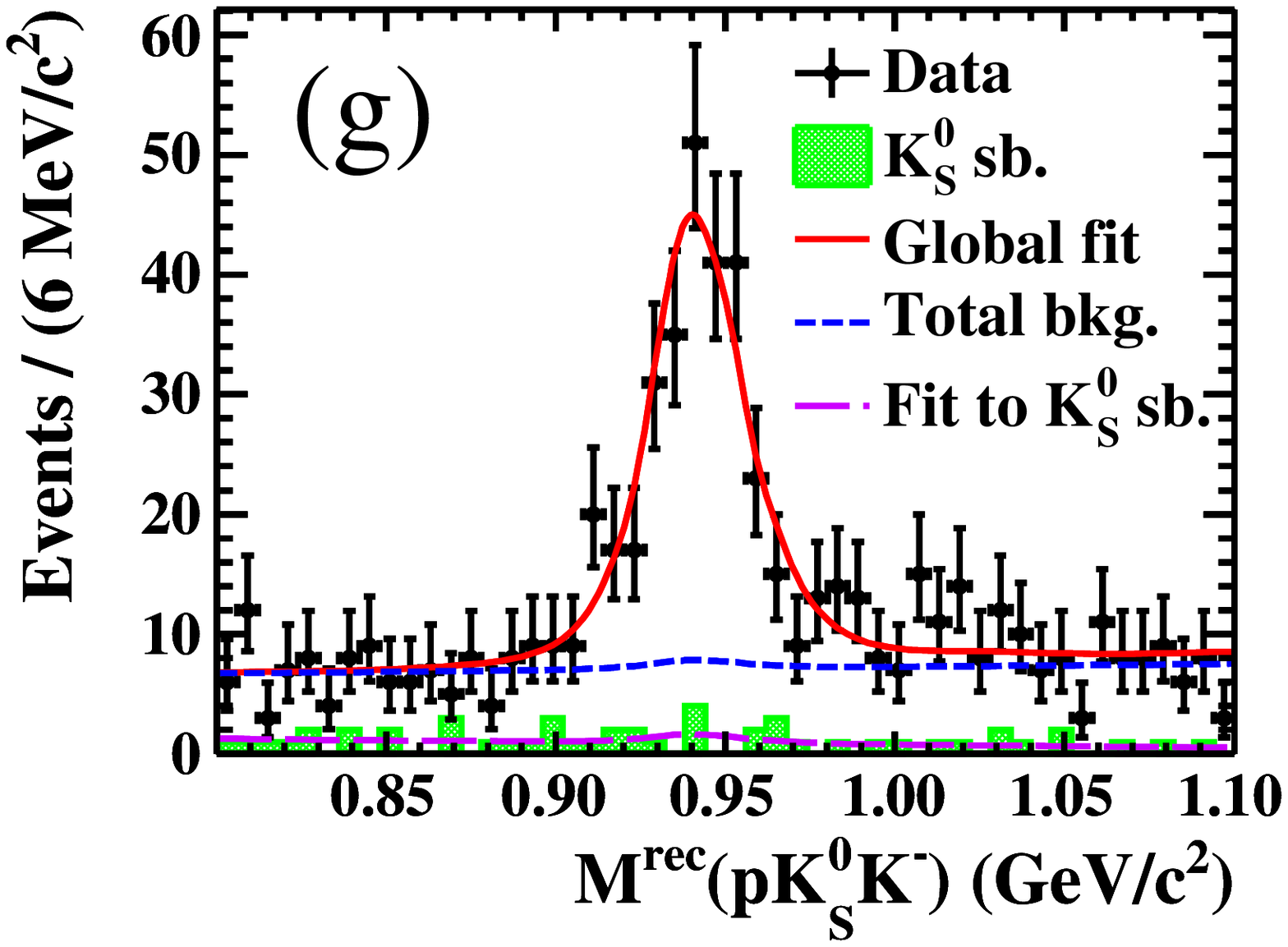}
    \end{overpic}    }
  {\caption{(Color online) Projections of the simultaneous fits to
      the $M^{\rm rec}(p\kshort K^-)$ spectra and $\kshort$
      mass side-band events in $\ee\rightarrow~p\kshort\bar{n}K^{-}$ at
      c.m.\ energies of (a) 3.773, (b) 4.008, (c) 4.226, (d) 4.258, (e)
      4.358, (f) 4.416, and (g) 4.600 GeV.  The dots with uncertainty bars
      are the signal candidate events in data, and the green-shaded
      histograms are shown as the normalized $\kshort$ mass side-band
      events in data.  The red solid curves show the total fits, the
      blue dashed lines are the total background components of the
      fits, and the violet long dashed curves are the fits to
      $\kshort$ mass side-band events.  }
    \label{fitneu}}
\end{figure*}

If multiple $pK^-$ combinations are found, the $\bar{n}$ candidate whose
$M^{\rm rec}(p\kshort K^-)$  is closest to the world
average anti-neutron mass value~\cite{PDG} is selected.  After the
above event selection criteria applied, a clear $\bar{n}$ signal
is observed in the $M^{\rm rec}(p\kshort K^-)$ at each c.m.\ energy, as
shown in Fig.~\ref{fitneu}.

To investigate non-$\kshort$ backgrounds, the $\kshort$ mass side-band
regions are selected as $0.4676<M_{\pi^+\pi^-}<0.4776$ GeV/$c^2$ or
$0.5176<M_{\pi^+\pi^-}<0.5276$ GeV/$c^2$.
According to the analysis of the inclusive MC samples
in the $K^0_S$ mass side-band regions, the main backgrounds
are from many processes with the $pK^-\pi^+\pi^-\bar{n}$ final state
with one weakly decaying hyperon like $\Lambda$ or $\Sigma$ involved,
where a small peak exists, as shown in Fig.~\ref{fitneu} in the shaded histograms.
Other background events, which form a smooth distribution
in the $M^{\rm rec}(pK^-K_S^0)$ spectra around 0.94 GeV/c$^2$,
are from numerous other processes, but none of them is dominant.

At each c.m.\ energy, an unbinned  maximum likelihood fit to
the $M^{\rm rec}(p\kshort K^-)$ spectra is performed to determine
the signal and background yields in the selected candidates within
[0.80, 1.10] GeV/c$^2$ and the normalized $\kshort$ mass side-band events.
The $\bar{n}$ signal shape is obtained through the
MC simulation at each c.m.\ energy smeared with a Gaussian function to
account for the difference in the resolution between the data and the
MC simulation.  The same $\bar{n}$ line shape is used for $\kshort$ mass
side-band events, and the other, non-peaking background contribution is described by a
first-order polynomial function.  Another first-order polynomial function associated
with the function of the fit to $\kshort$ side-bands represents the
remaining background contribution.  The parameters of the Gaussian
function and the two first-order polynomials are left free.  The fits
of the $M^{\rm rec}(p\kshort K^-)$ spectra at the seven
c.m.\ energies are shown in Fig.~\ref{fitneu}, and the signal yields
along with other numerical results are summarized in
Table~\ref{tab:neu}.

\begin{table*}[htbp]
  \caption{The c.m.\ energy $(\sqrt{s})$, integrated luminosity $(\mathcal{L})$,  detection efficiency $(\epsilon)$,
    vacuum polarization ($\frac{1}{|1-\Pi|^2}$) and radiative correction factor $(1+\delta)$, number of fitted $\bar{n}$ signal events $(N_{\rm sig})$ excluding the peaking backgrounds,
    and Born cross section ($\sigma_{\rm B}$) of $e^{+}e^{-}\rightarrow pK^{0}_{S}\bar{n}K^{-}$ at each energy point. The first uncertainties are statistical and the second systematic.}
\label{tab:neu}
\centering
\begin{tabular}{C{2cm}|C{2cm}|C{2cm}|C{2cm}|C{2cm}|C{3.1cm}|C{3.8cm}}
  \hline
  \hline
  $\sqrt{s}~$(GeV) & $\mathcal{L}~(\rm pb^{-1})$ & $\epsilon~(\%)$ & $\frac{1}{|1-\Pi|^2}$ & $1+\delta$ & $N_{\rm sig}$
  & $\sigma_{\rm B}~(\rm pb)$\\
  \hline
  $3.773$ & $2931.8$ & $33.44$ & $1.057$ &  $0.881$  &  $2317\pm{62}$ &  $3.67\pm0.10\pm0.20$  \\
  $4.008$ & $482.0$  & $35.30$ & $1.044$ &  $0.926$  &  $367\pm{24}$  &  $3.22\pm0.21\pm0.17$  \\
  $4.226$ & $1047.3$ & $37.03$ & $1.056$ &  $0.934$  &  $718\pm{38}$  &  $2.71\pm0.14\pm0.14$  \\
  $4.258$ & $825.7$  & $37.26$ & $1.054$ &  $0.936$  &  $527\pm{30}$  &  $2.51\pm0.14\pm0.14$  \\
  $4.358$ & $539.8$  & $38.04$ & $1.051$ &  $0.952$  &  $325\pm{24}$  &  $2.29\pm0.17\pm0.12$  \\
  $4.416$ & $1028.9$ & $38.51$ & $1.053$ &  $0.960$  &  $563\pm{32}$  &  $2.03\pm0.12\pm0.12$  \\
  $4.600$ & $566.9$  & $39.91$ & $1.055$ &  $0.967$  &  $264\pm{23}$  &  $1.65\pm0.14\pm0.09$  \\

  \hline   			  	 	 	 	 	
  \hline
\end{tabular}
\end{table*}

\subsection{\boldmath $\Lambda(1520)$ modes}

In addition to the common selection criteria for $p$ or $K$
candidates, the combined TOF and $dE/dx$ information is used to
calculate $\chi^{2}_{\rm PID}(i)~(i=p,~K)$ for each hadron ($i$)
hypothesis, and a one-constraint (1C) kinematic fit is performed with
the $p,~K^{0}_{S},~K^{-}$, and $\bar{n}$ combination by constraining the
missing mass of the undetected $\bar{n}$ to its nominal
mass~\cite{PDG}. The combination with the minimum $\chi^2_{\rm
  sum}=\chi^{2}_{1C}+\chi^{2}_{\rm PID}(p)+\chi^{2}_{\rm PID}(K^-)$ in an event is
selected, and we require $\chi^2_{\rm sum}<20$, where $\chi^{2}_{1C}$
is the $\chi^{2}$ of the kinematic fit.

\begin{figure*}[htbp]
    \centering
    \hskip -0.4cm \mbox{
    \hskip -0.48cm
    \begin{overpic}[width=5.2cm,height=4.0cm,angle=0]{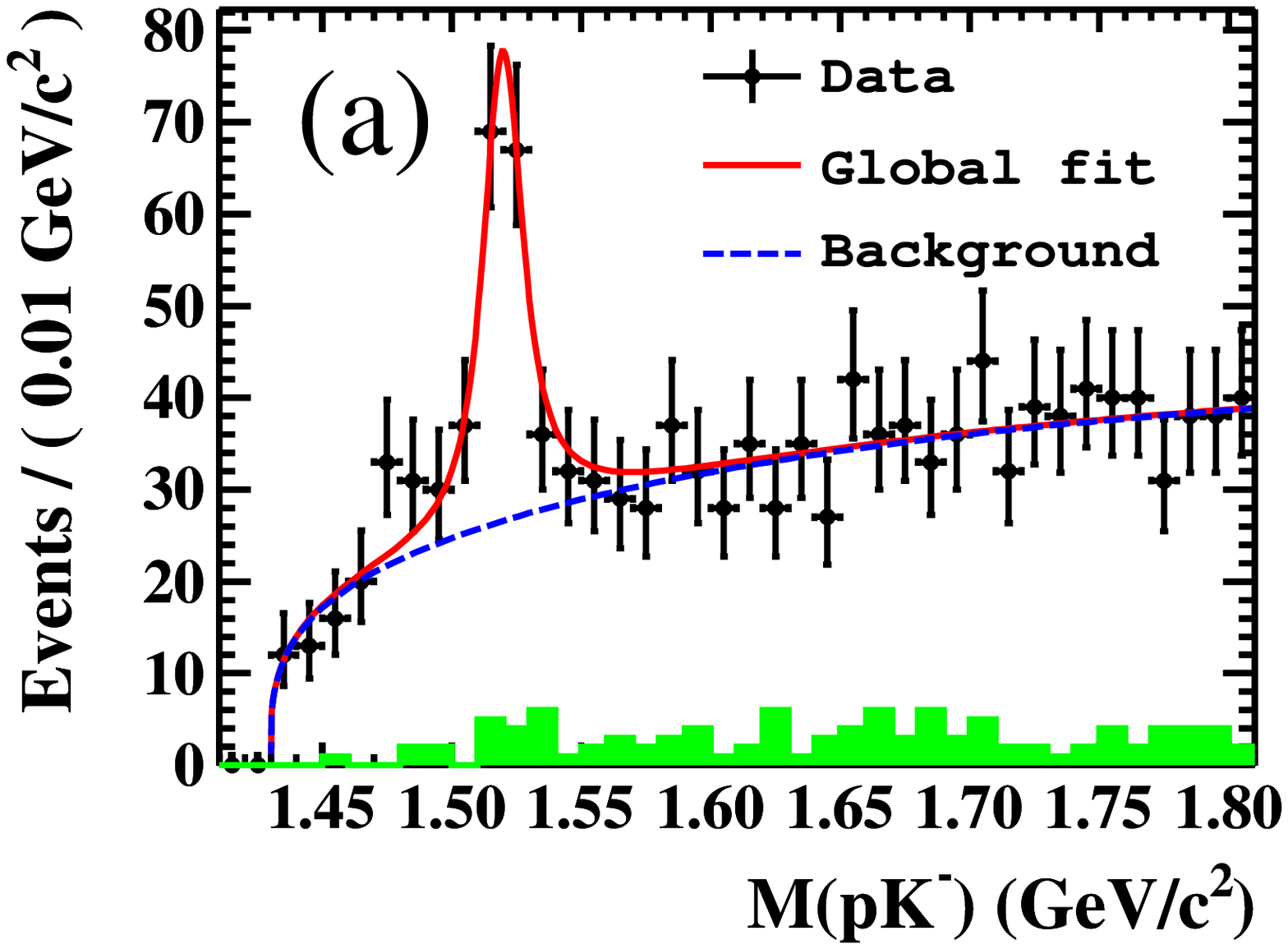}
    \end{overpic}
    \hskip -0.71cm
    \begin{overpic}[width=5.2cm,height=4.0cm,angle=0]{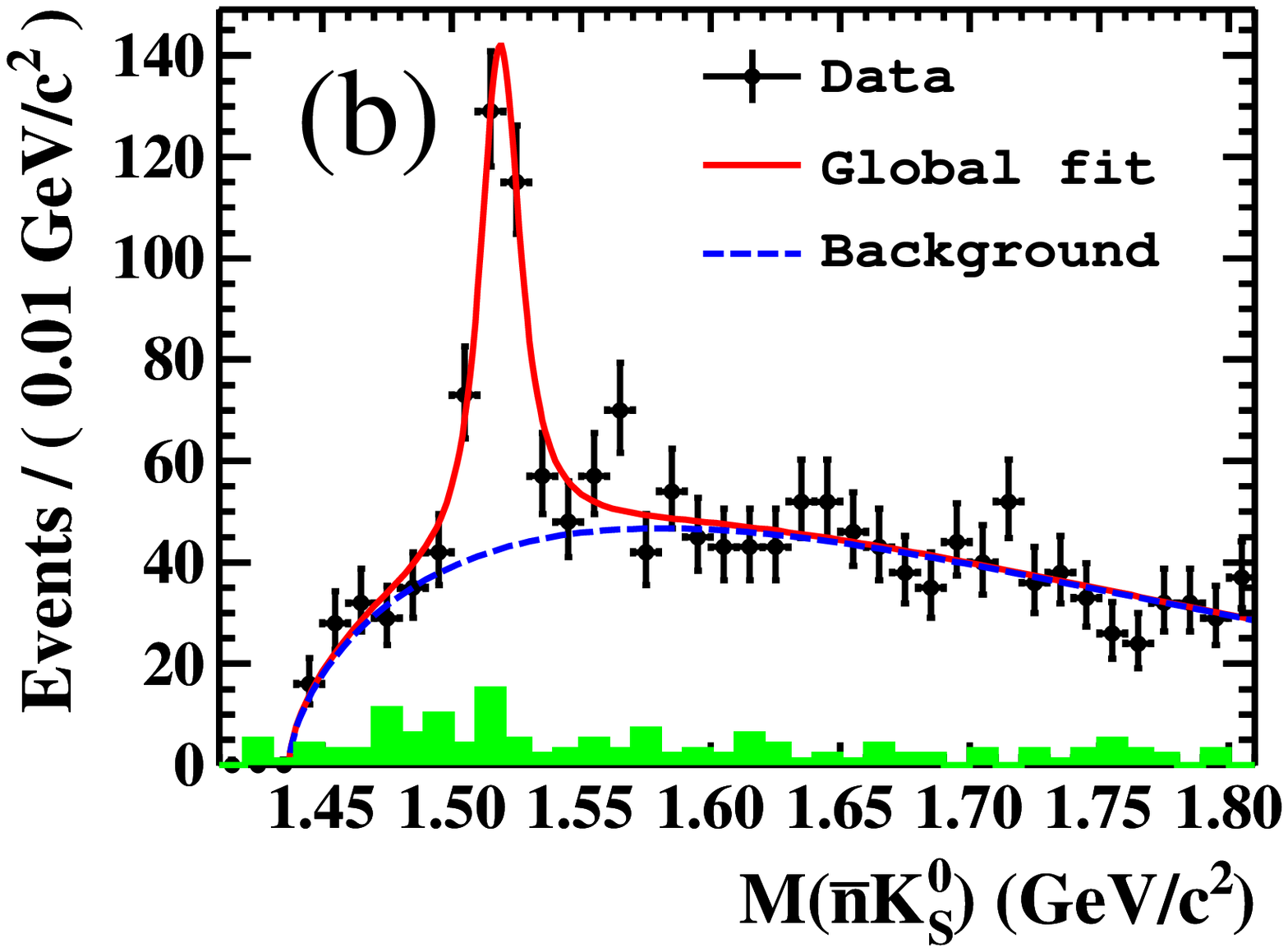}
    \end{overpic}
    \hskip -0.71cm
    \begin{overpic}[width=5.2cm,height=4.0cm,angle=0]{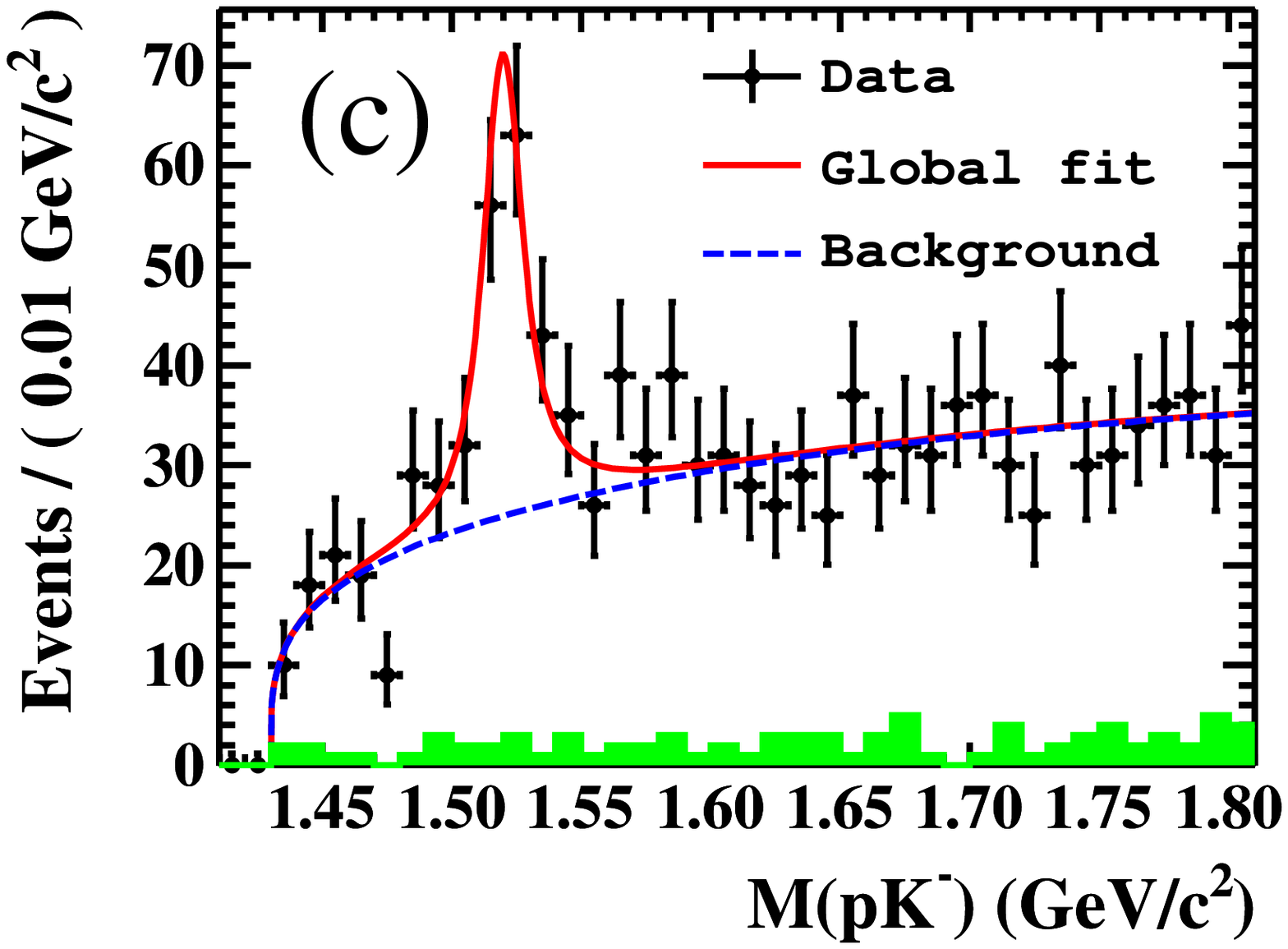}
    \end{overpic}
    \hskip -0.71cm
    \begin{overpic}[width=5.2cm,height=4.0cm,angle=0]{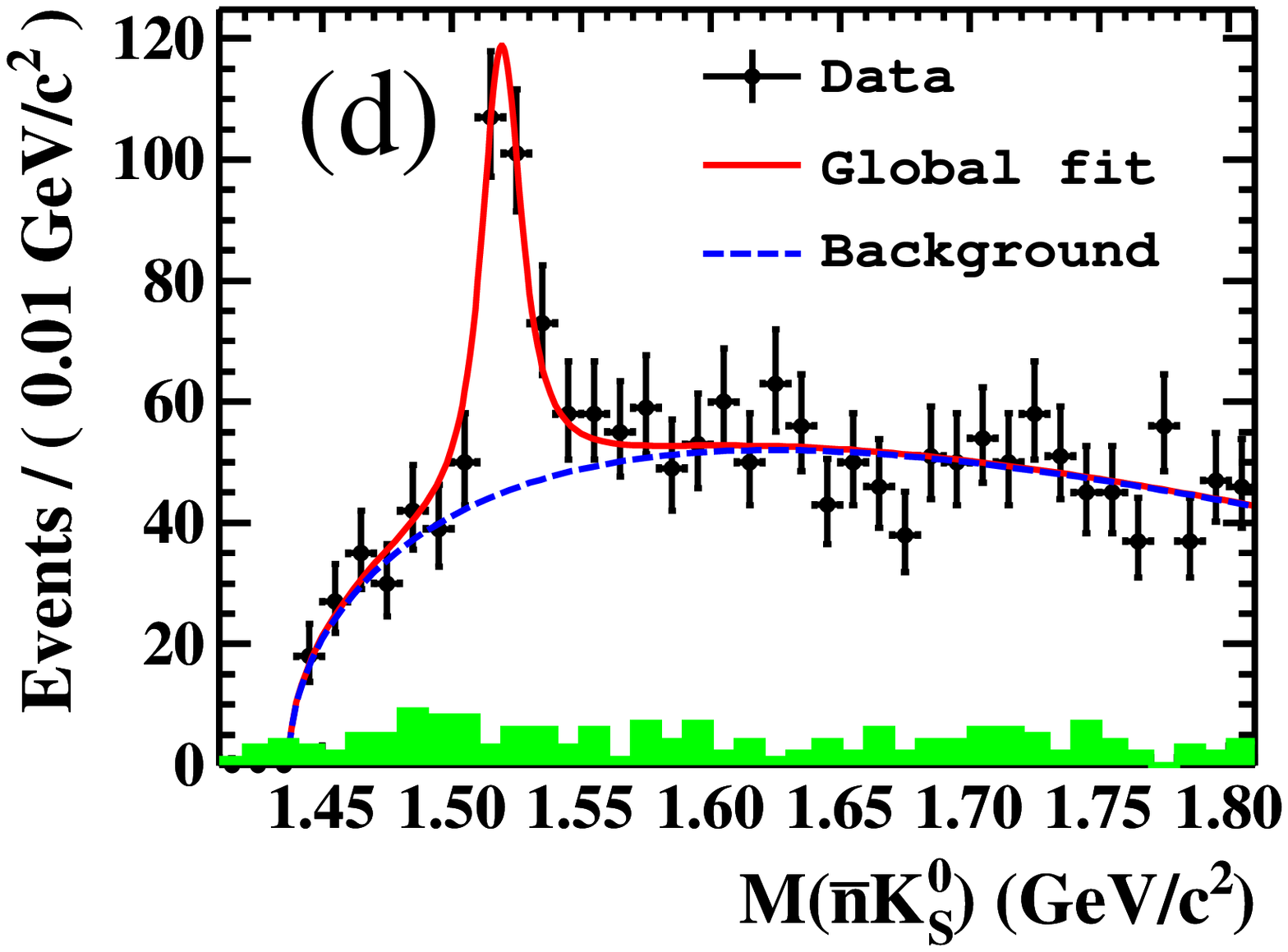}
    \end{overpic}
    }
    {\caption{ (Color online)
        Fits to the $pK^-$ and $\bar{n}K^{0}_{S}$ invariant
        mass distributions to determine signal yields for $\ee\to\Lambda(1520)\bar{n}K^0_S$
        and $\ee\to pK^-\bar{\Lambda}(1520)$, respectively,
        where (a) and (b) are from the data at the c.m.\ energy of 3.773 GeV,
        and (c) and (d) are from the full $XYZ$ data.  Clear
        $\Lambda(1520)$ signals are observed.  The red solid lines
        show the global fits, and the blue dashed curves show the total fitted backgrounds
        with and without $pK^-\bar{n}K^{0}_{S}$ final states.
        The green-shaded histograms are the contributions of the normalized $K^0_S$ mass side-band events in data.}
     \label{fitL1520}}
\end{figure*}

After the event selection requirements have been applied, clear
$\Lambda(1520)$ signals are found for the processes
$\ee\rightarrow\Lambda(1520)\bar{n}K^{0}_{S}$ and
$\ee\rightarrow\Lambda(1520)\bar{p}K^{+}$ in data at the c.m.\ energy of
 3.773 GeV and in the full $XYZ$ data, as shown in
Fig.~\ref{fitL1520}.

\begin{figure*}[htbp]
    \centering
    \hskip -0.4cm \mbox{
    \hskip -0.4cm
    \begin{overpic}[scale=0.26]{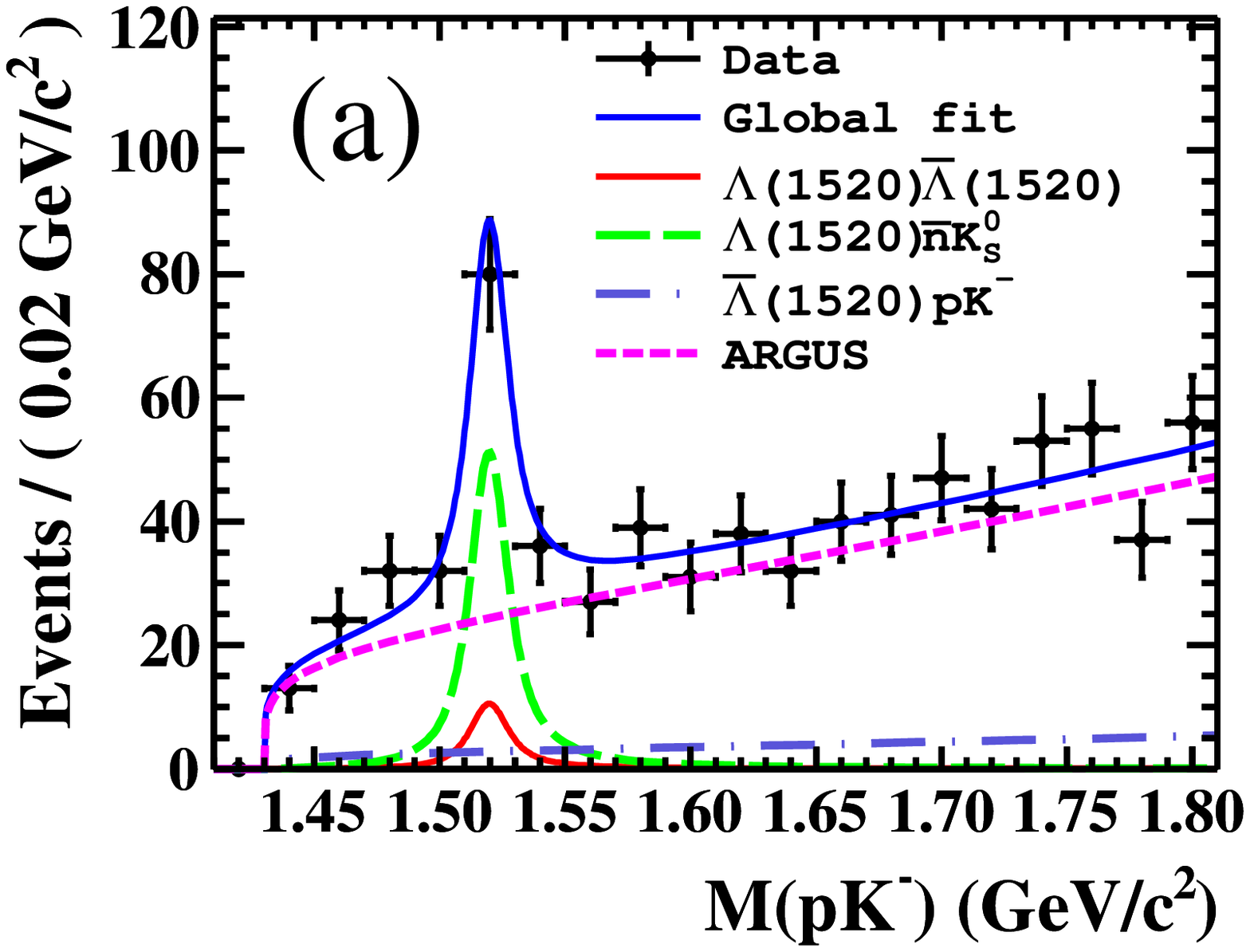}
    \end{overpic}
    \hskip -0.75cm
    \begin{overpic}[scale=0.26]{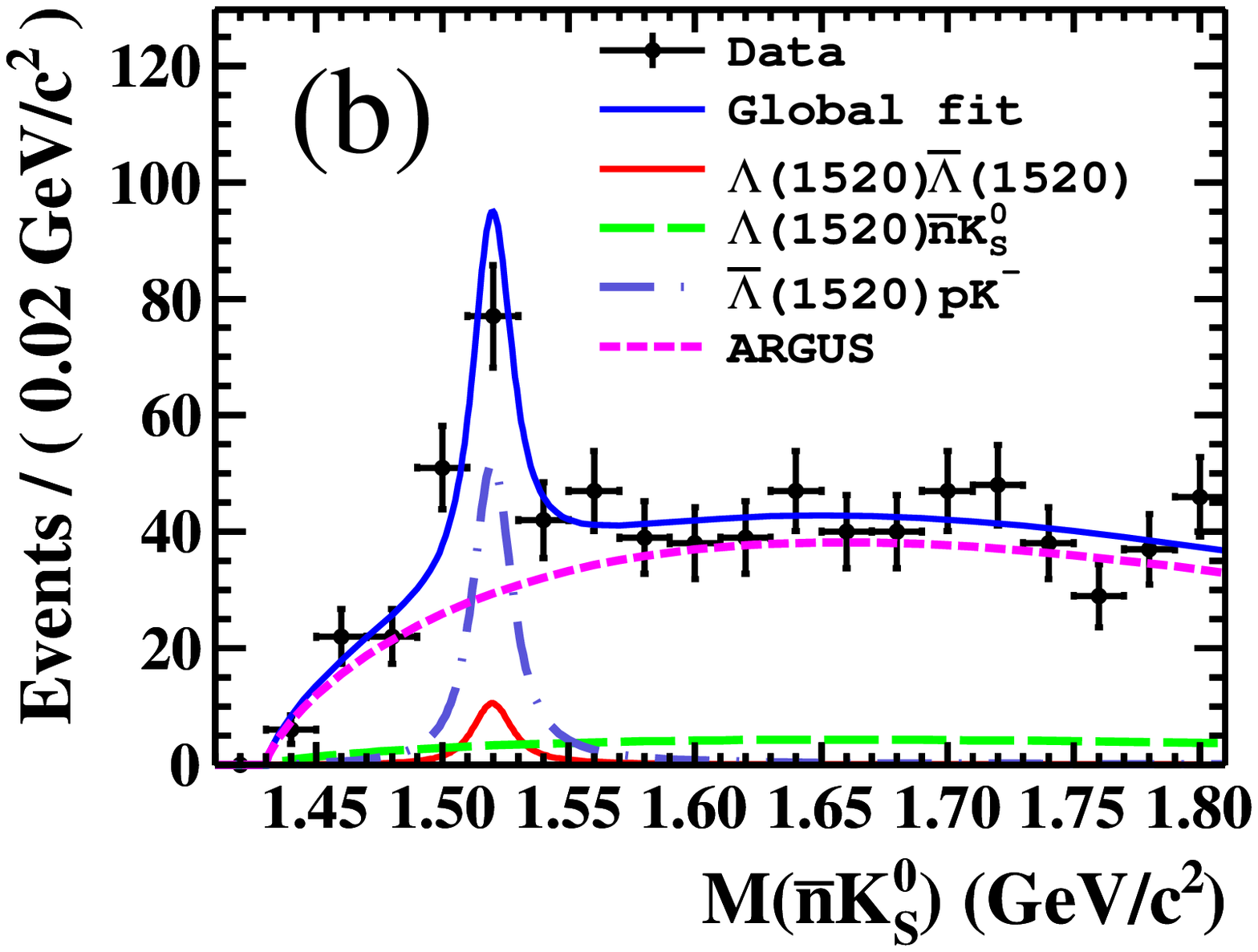}
    \end{overpic}
    \hskip -0.75cm
    \begin{overpic}[scale=0.26]{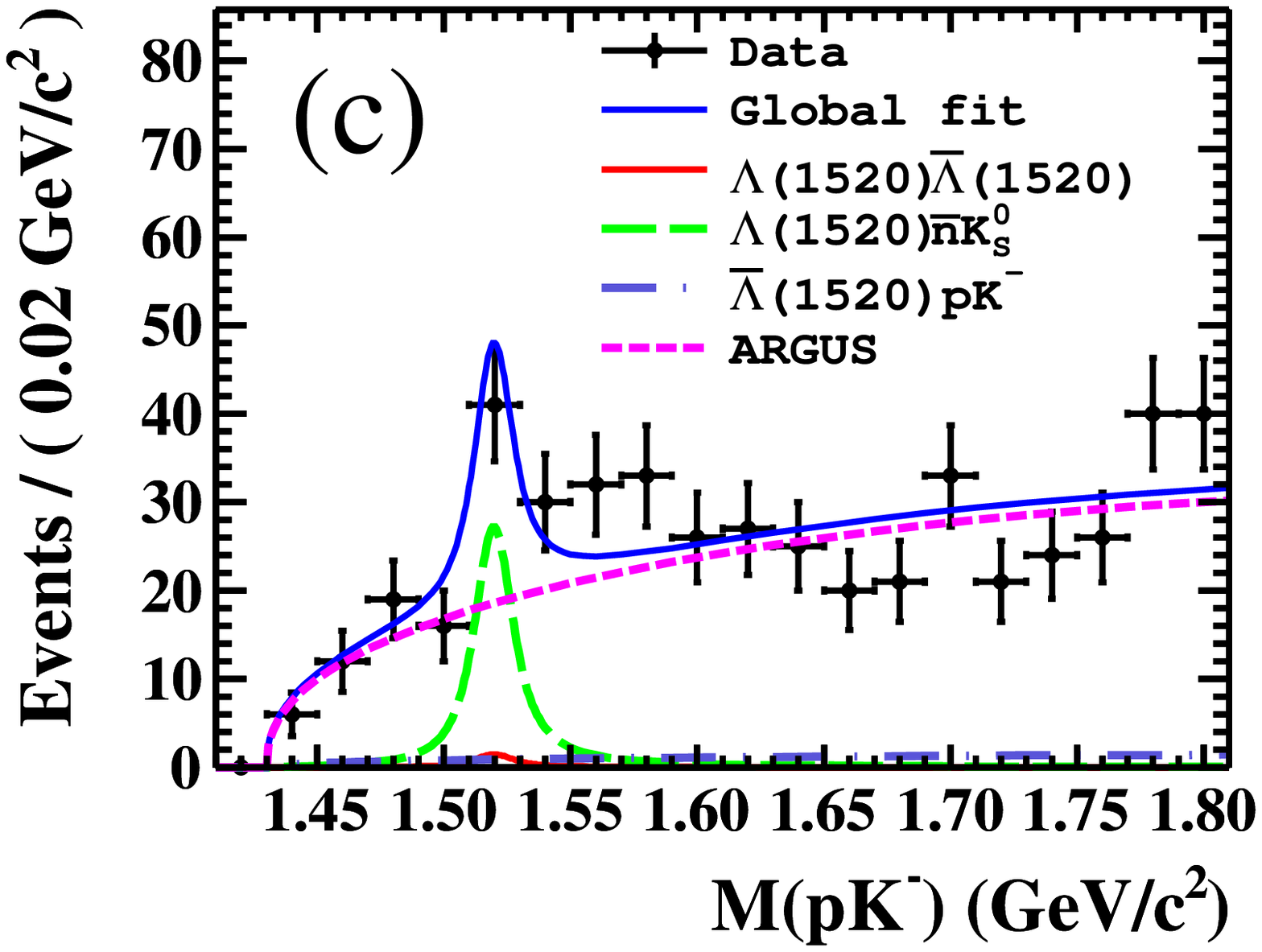}
    \end{overpic}
    \hskip -0.75cm
    \begin{overpic}[scale=0.26]{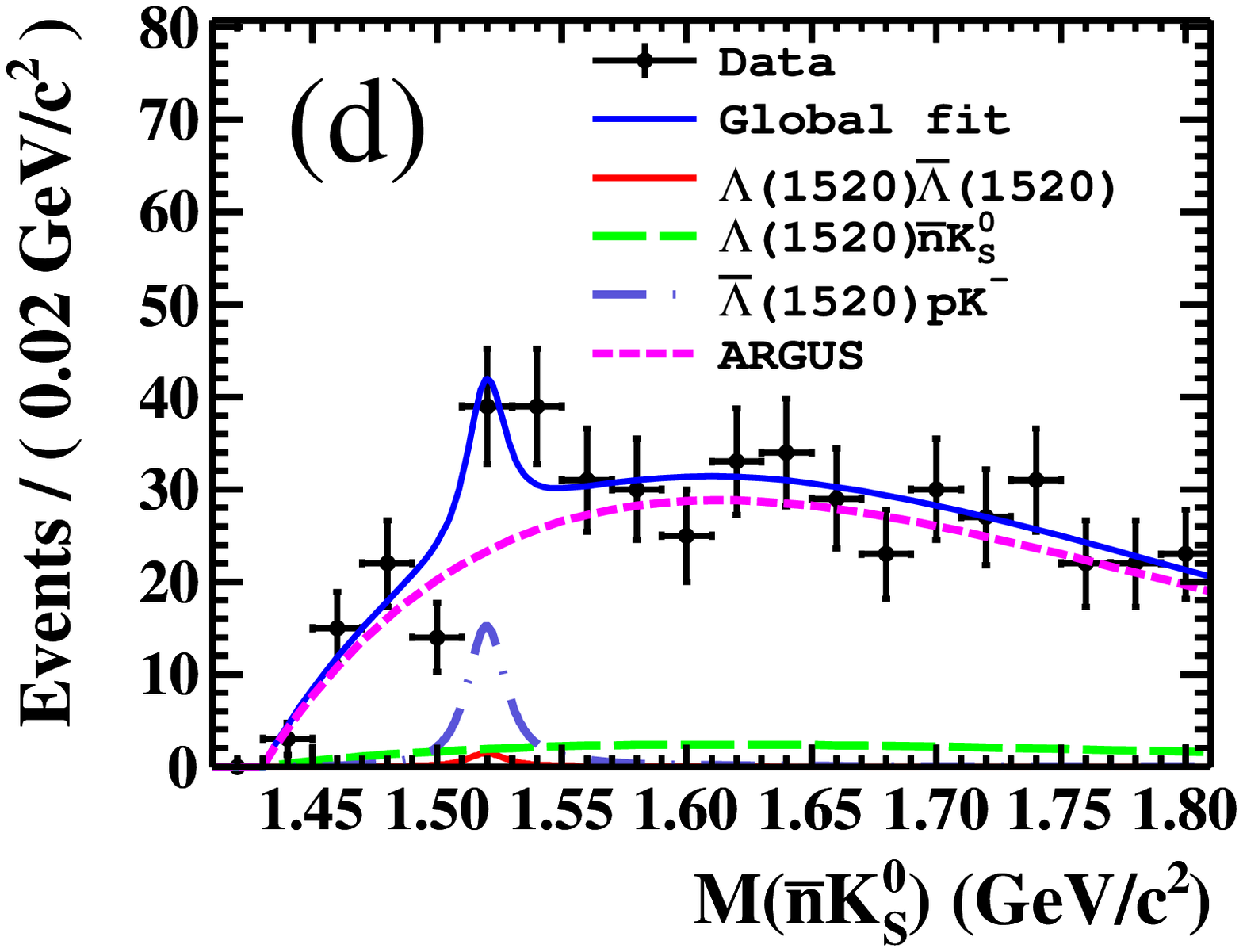}
    \end{overpic}
    \hskip -0.3cm
    }
    {\caption{(Color online) Projections of the 2D fits described in
        the text to the $M(pK^{-})$ and $M(\bar{n}K^{0}_{S})$
        distributions, where (a) and (b) are from the data at
        the c.m.\ energy of 3.773 GeV, and (c) and (d) are from the full
        $XYZ$ data.  The blue solid lines show the best fits, the red
        solid lines show the
        $\ee\rightarrow\Lambda(1520)\bar{\Lambda}(1520)$ signals,
        the pink dashed lines represent the fitted non-resonant backgrounds,
        the gray-blue long dot-dashed lines show the contributions
        from $\ee\rightarrow\Lambda(1520)\bar{p}K^{+}$, and the
        green long dashed lines indicate the contributions from
        $\ee\rightarrow\Lambda(1520)\bar{n}K^{0}_{S}$.  }
     \label{fit2D-L1520}}
\end{figure*}

According to the analysis of the inclusive MC samples, the dominant
background events are from processes with $pK^{0}_{S}\bar{n}K^-$ final
states without a $\Lambda(1520)$.  The remaining backgrounds are
from many processes with only a few events in each mode.  No
peaking background is found in the $pK^-$ or $nK^{0}_{S}$ invariant mass
spectra at around 1520~MeV/$c^2$.

An unbinned maximum likelihood fit is performed to the $pK^{-}$ and $nK^{0}_{S}$
invariant mass distributions to determine the $\Lambda(1520)$ yields individually.
The $\Lambda(1520)$ signal is described by a D-wave relativistic Breit-Wigner (BW) function
with an energy-dependent width $\Gamma_{pK}$ convolved with a Gaussian
function.  The $\Lambda(1520)$ mass and width are fixed to the
world average values~\cite{PDG}.  The $pK^{-}$ and $n K^{0}_{S}$ mass
resolutions of the signal Gaussian functions are the same and fixed to
3~MeV/$c^2$ as determined from fits to the individual MC
distributions.  The background shape is parameterized by an ARGUS
function~\cite{ref:argus}.

For $\Lambda(1520)\to pK^{-}$ (the same for $\Lambda(1520)\to
n K^{0}_{S}$), $\Gamma_{pK}$~\cite{width-dependent} is described by:
\begin{equation}\label{width-equ}
   \Gamma_{pK}=\Gamma_{r}\left(\frac{P_{pK}}{P_{r}}\right)^{2L+1}\left(\frac{M_{r}}{M_{pK}}\right)F_{r}^{2},
\end{equation}
where $M_{pK}$ is the $pK$ invariant mass, $M_r$ the $\Lambda(1520)$ nominal mass~\cite{PDG},
$P_{pK}$ the momentum of the daughter particle $p$ or $K$ in the $pK$ rest frame
($P_r$ when the $pK$ invariant mass is the $\Lambda(1520)$ nominal mass~\cite{PDG}),
$L$ the $\Lambda(1520)$ decay orbital angular momentum, $\Gamma_r$ the
$\Lambda(1520)$ nominal width~\cite{PDG}, and $F_r$ the Blatt-Weisskopf penetration form factor.
For $\Lambda(1520)\to pK^{-}$ and $n K^{0}_{S}$ decays ($L=2$), $
F_{r}=\sqrt{9+3R^{2}P_{r}^{2}+R^{4}P_{r}^{4}}/\sqrt{9+3R^{2}P_{pK}^{2}+R^{4}P_{pK}^{4}},
$ where $R$ is a phenomenological factor
with little sensitivity to the $\Gamma_{pK}$ and generally taken as
$R=5~\hbox{GeV}^{-1}$~\cite{width-dependent}.

To avoid double counting from
$e^+e^-\rightarrow\Lambda(1520)\bar{\Lambda}(1520)$ in
calculating the Born cross sections of
$e^+e^-\rightarrow\Lambda(1520)\bar{n}K^{0}_{S}$ and
$\Lambda(1520)\bar{p}K^{+}$,
$\Lambda(1520)\bar{\Lambda}(1520)$ pair events are subtracted
from the three-body decays,
since it is difficult to perform a two-dimensional (2D) fit in the full range.
The final number of the signal events and
the corresponding statistical significance of the
$e^+e^-\rightarrow\Lambda(1520)\bar{n}K^{0}_{S}$ and
$\Lambda(1520)\bar{p}K^{+}$ signal at each c.m.\ energy are listed in
Table~\ref{tab:L1520}. The statistical significance is calculated
using $\sqrt{-2\ln(\mathcal{L}_0/\mathcal{L}_{\rm max})}$, where
$\mathcal{L}_{\rm max}$ and $\mathcal{L}_0$ are the likelihoods of the
fits with and without the $\Lambda(1520)$ signal included, respectively.

We extend the unbinned maximum likelihood fit described above into a
2D fit to the $pK^{-}$ versus $\bar{n}K^{0}_{S}$ mass spectra
to determine the yield of the process $e^+e^-\rightarrow\Lambda(1520)\bar{\Lambda}(1520) \to
pK^{-}\bar{n}K^{0}_{S}$.  We assume that the two discriminating variables $M(pK^-)$ and $M(nK_S^0)$
are uncorrelated, and the 2D probability density function (PDF) is the product
of two one-dimensional (1D) PDFs for the two variables.
The total PDFs include four components: $\Lambda(1520)\bar{\Lambda}(1520)$,
$\Lambda(1520)\bar{n}K_S^0$, $pK^-\bar{\Lambda}(1520)$, and non$-\Lambda(1520)$.
The signal shapes of $\Lambda(1520)(\rightarrow pK^-)$ and
$\bar{\Lambda}(1520)(\rightarrow \bar{n}K^0_S)$ in the 2D fit
are the same as those in $e^+e^-\rightarrow\Lambda(1520)\bar{n}K^{0}_{S}$ and
$\Lambda(1520)\bar{p}K^{+}$, respectively. All of the backgrounds are parameterized
by an ARGUS function~\cite{ref:argus}.
The projections of the 2D fits in data
at the c.m.\ energy of 3.773 GeV and in the full $XYZ$ data are shown in
Figs.~\ref{fit2D-L1520}(a-d).

Since only a few $\Lambda(1520)\bar{\Lambda}(1520)$
pair signal events are observed and the statistical significance
is less than 3.0 standard deviations ($\sigma$) at each c.m.\ energy,
the upper limit on the number of signal events, $N_{\rm sig}^{\rm
  up}$, is determined at the $90\%$ confidence level (C.L.) by solving
\begin{equation}\label{ul-equ}
  \int^{N_{\rm sig}^{\rm up}}_0 \mathcal{L}(x)dx / \int^{+\infty}_0\mathcal{L}(x)dx = 0.9,
\end{equation}
where $x$ is the number of fitted signal events, and $\mathcal{L}(x)$ is the
likelihood function in the fit to data.

To account for the systematic uncertainties, the likelihood distribution
is convolved with a Gaussian function $G(x; 0, \sigma)$ with a standard deviation of
$\sigma=x\times\Delta$,
\begin{equation}\label{Lsigma}
  \mathcal{L}^{\prime}(\mu) = \int^{+\infty}_0 \mathcal{L}(x)\times G(\mu - x; 0, \sigma) dx,
\end{equation}
where $\mu$ is the expected number of signal events, $\mathcal{L}^{\prime}(\mu)$ indicates the expected likelihood distribution,
and $\Delta$ refers to the total relative systematic uncertainty discussed in Section~\ref{section7}.
The upper limit on the number of $\Lambda(1520)\bar{\Lambda}(1520)$ pair events
and statistical significance at each energy are listed in Table~\ref{tab:L1520}.

\begin{table*}[htbp]
  \caption{The c.m.\ energy $(\sqrt{s})$, integrated luminosity $(\mathcal{L})$,  detection efficiency $(\epsilon$), vacuum polarization ($\frac{1}{|1-\Pi|^2}$), radiative correction factor $(1+\delta)$, number of observed signal events $(N_{\rm sig})$, statistical signal significance ($S$), and
    calculated (90\% C.L. upper limit of) Born cross section ($\sigma_{\rm B}$) are listed for the studied $\Lambda(1520)$ modes at each energy point.
    The first uncertainties are statistical and the second systematic.
    For the 90\% C.L. upper limits, the systematic uncertainties have been included.}
\label{tab:L1520}
\centering
\begin{tabular*}{\textwidth}{@{\extracolsep{\fill}}c|cccccccc}
  \hline
  \hline
  Mode & $\sqrt{s}$~(GeV) & $\mathcal{L}~(\rm pb^{-1})$ & $\epsilon~(\%)$ & $\frac{1}{|1-\Pi|^2}$ & $1+\delta$ &  $N_{\rm sig}$ &
  $S~(\sigma)$ &$\sigma_{\rm B}~(\rm pb)$\\
  \hline
  \multirow{6}*{ \begin{sideways} $\Lambda(1520)\bar{n}K_{S}^0$ \end{sideways} }
              & $3.773$ & $2931.8$ & $23.87$ & $1.057$ &  $0.878$  &  $122\pm{21}$    & 8.3 &  $1.21\pm0.21\pm0.09$  \\
              & $4.008$ & $482.0$  & $24.64$ & $1.044$ &  $0.927$  &  $24.7\pm{9.0}$  & 3.5 &  $1.38\pm0.50\pm0.10$  \\
              & $4.226$ & $1047.3$ & $25.34$ & $1.056$ &  $0.933$  &  $20.5\pm{9.4}$  & 3.1 &  $0.50\pm0.23\pm0.04$  \\
              & $4.258$ & $825.7$  & $25.44$ & $1.054$ &  $0.936$  &  $21.0\pm{7.8}$  & 3.3 &  $0.65\pm0.24\pm0.04$  \\
              & $4.358$ & $539.8$  & $25.76$ & $1.051$ &  $0.954$  &  $8.3\pm{5.9}$   & 3.0 &  $0.38\pm0.27\pm0.03$  \\
              & $4.416$ & $1028.9$ & $25.95$ & $1.053$ &  $0.962$  &  $25.5\pm{8.7}$  & 4.0 &  $0.61\pm0.21\pm0.04$  \\
              & $4.600$ & $566.9$  & $26.53$ & $1.055$ &  $0.970$  &  $10.3\pm{6.1}$  & 4.0 &  $0.43\pm0.25\pm0.03$  \\
  \hline
  \multirow{6}*{ \begin{turn}{90} $\Lambda(1520)\bar{p}K^{+}$ \end{turn} }
              & $3.773$ & $2931.8$ & $27.22$ & $1.057$ &  $0.879$  &  $250\pm{27}$    & 11.9 & $4.33\pm0.47\pm0.28$  \\
              & $4.008$ & $482.0$  & $27.33$ & $1.044$ &  $0.931$  &  $40\pm{11}$     & 4.3  & $4.01\pm1.10\pm0.27$  \\
              & $4.226$ & $1047.3$ & $27.45$ & $1.056$ &  $0.935$  &  $60\pm{14}$     & 5.6  & $2.72\pm0.63\pm0.18$  \\
              & $4.258$ & $825.7$  & $27.46$ & $1.054$ &  $0.936$  &  $24.9\pm{8.7 }$ & 3.9  & $1.43\pm0.50\pm0.10$  \\
              & $4.358$ & $539.8$  & $27.51$ & $1.051$ &  $0.951$  &  $16.1\pm{8.1 }$ & 3.1  & $1.39\pm0.70\pm0.10$  \\
              & $4.416$ & $1028.9$ & $27.54$ & $1.053$ &  $0.957$  &  $46\pm{12}$     & 4.5  & $2.07\pm0.54\pm0.14$  \\
              & $4.600$ & $566.9$  & $27.63$ & $1.055$ &  $0.974$  &  $6.4 \pm{6.8 }$ & 3.0  & $0.51\pm0.54\pm0.04$  \\
  \hline
  \multirow{6}*{ \begin{sideways} $\Lambda(1520)\bar{\Lambda}(1520)$ \end{sideways}}
              & $3.773$ & $2931.8$ & $27.65$ & $1.057$ &  $0.882$  &  $<24~(13.9\pm7.5)$    & 2.1 &  $<1.9$    \\
              & $4.008$ & $482.0$  & $28.77$ & $1.044$ &  $0.928$  &  $<5.5~(0.0\pm3.5)$    & 0.1 &  $<2.4$    \\
              & $4.226$ & $1047.3$ & $29.81$ & $1.056$ &  $0.932$  &  $<7.5~(1.6\pm3.8)$    & 0.5 &  $<1.4$    \\
              & $4.258$ & $825.7$  & $29.95$ & $1.054$ &  $0.939$  &  $<7.7~(2.4\pm2.0)$    & 1.6 &  $<1.8$    \\
              & $4.358$ & $539.8$  & $30.42$ & $1.051$ &  $0.954$  &  $<2.8~(0.0\pm0.8)$    & 0.3 &  $<1.0$    \\
              & $4.416$ & $1028.9$ & $30.71$ & $1.053$ &  $0.956$  &  $<5.3~(0.3\pm2.9)$    & 0.1 &  $<1.0$    \\
              & $4.600$ & $566.9$  & $31.55$ & $1.055$ &  $0.970$  &  $<2.4~(0.0\pm0.8)$    & 0.1 &  $<0.8$    \\
  \hline
  \hline
\end{tabular*}
\end{table*}

\begin{figure}[htbp]
    \centering
    \hskip -0.4cm \mbox{
    \hskip -0.4cm
    \begin{overpic}[scale=0.25,angle=0]{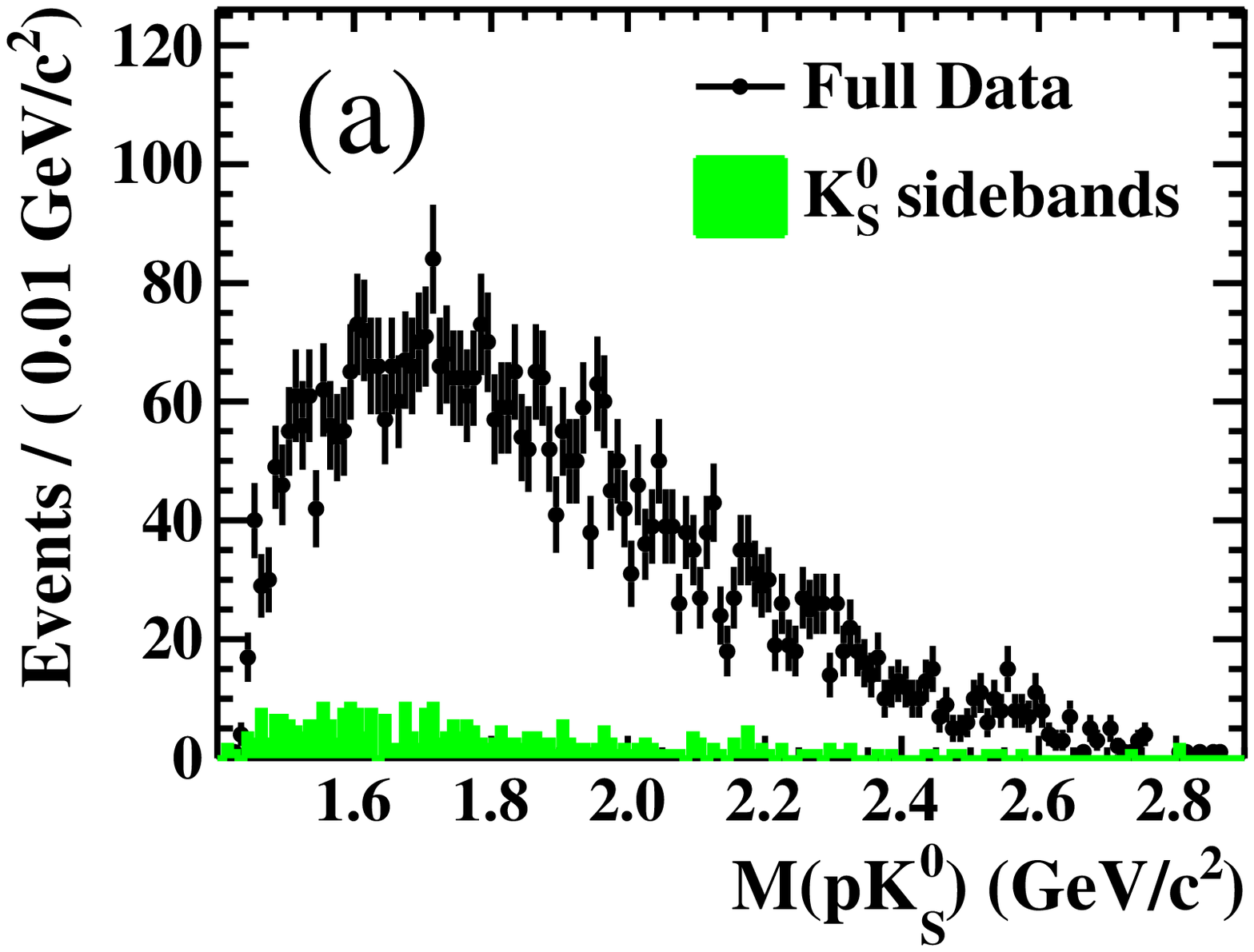}
    \end{overpic}
    \hskip -0.75cm
    \begin{overpic}[scale=0.25,angle=0]{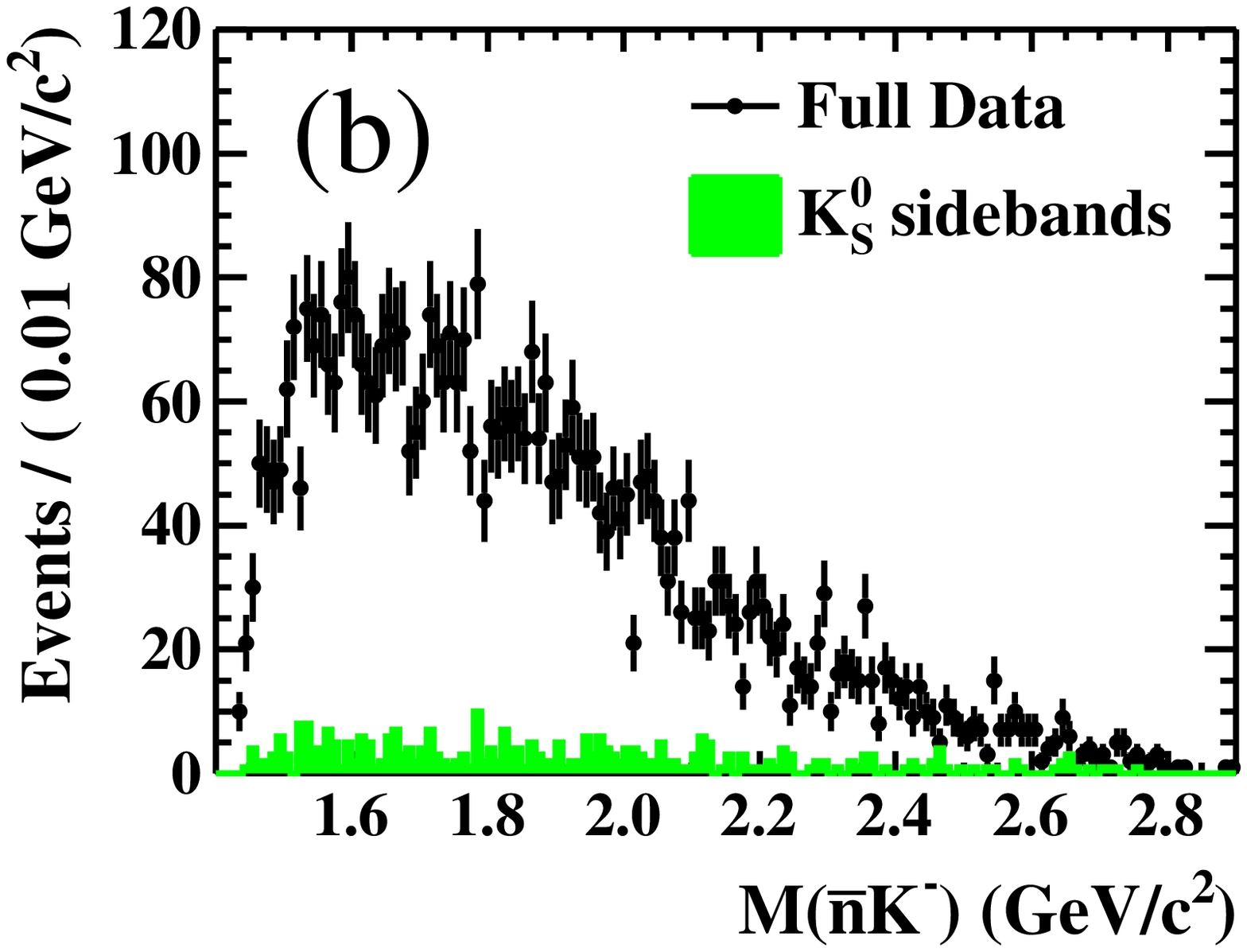}
    \end{overpic}}
    \vskip -0.2cm
    \hskip -0.0cm \mbox{
    \hskip -0.4cm
    \begin{overpic}[scale=0.25,angle=0]{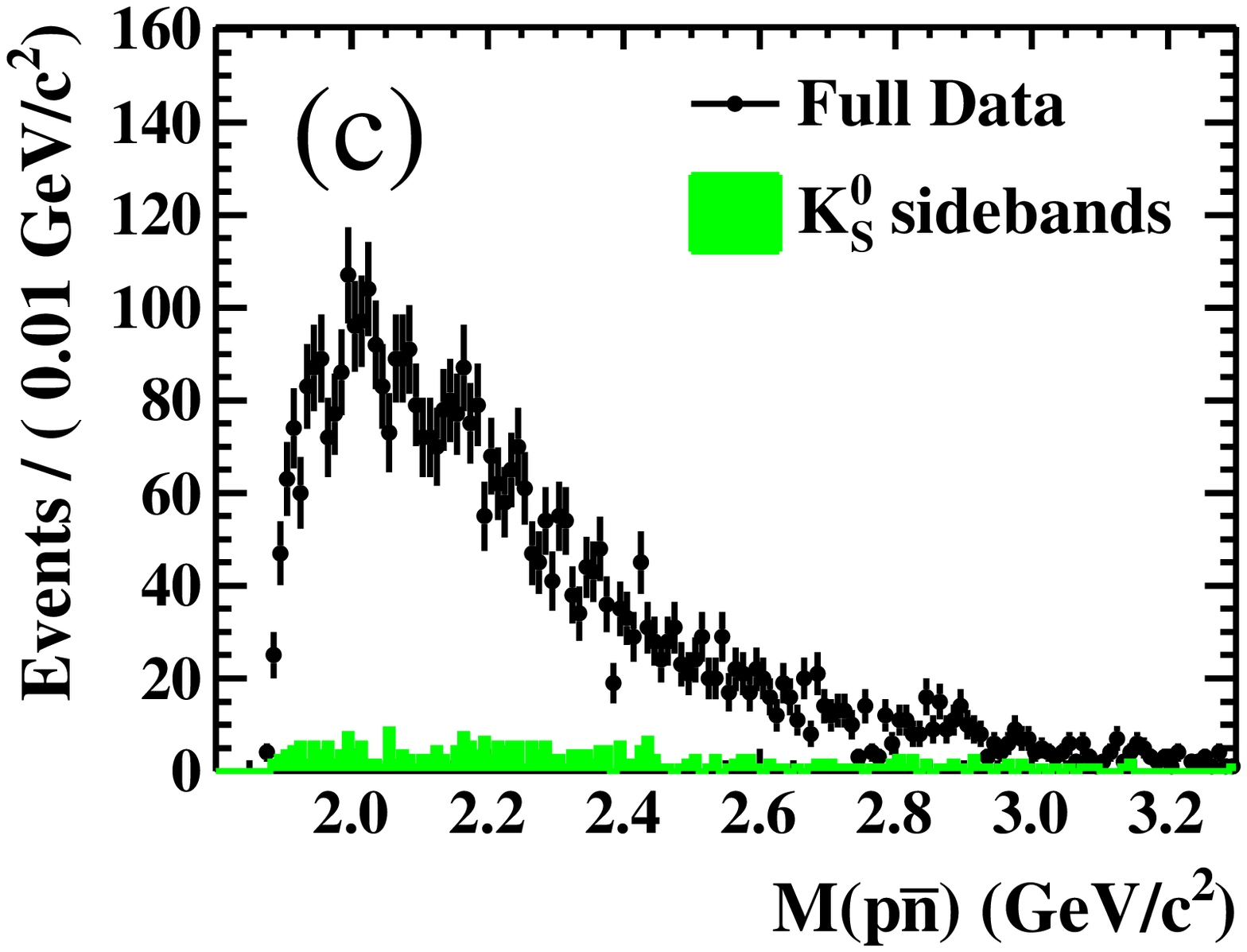}
    \end{overpic}
    \hskip -0.75cm
    \begin{overpic}[scale=0.25,angle=0]{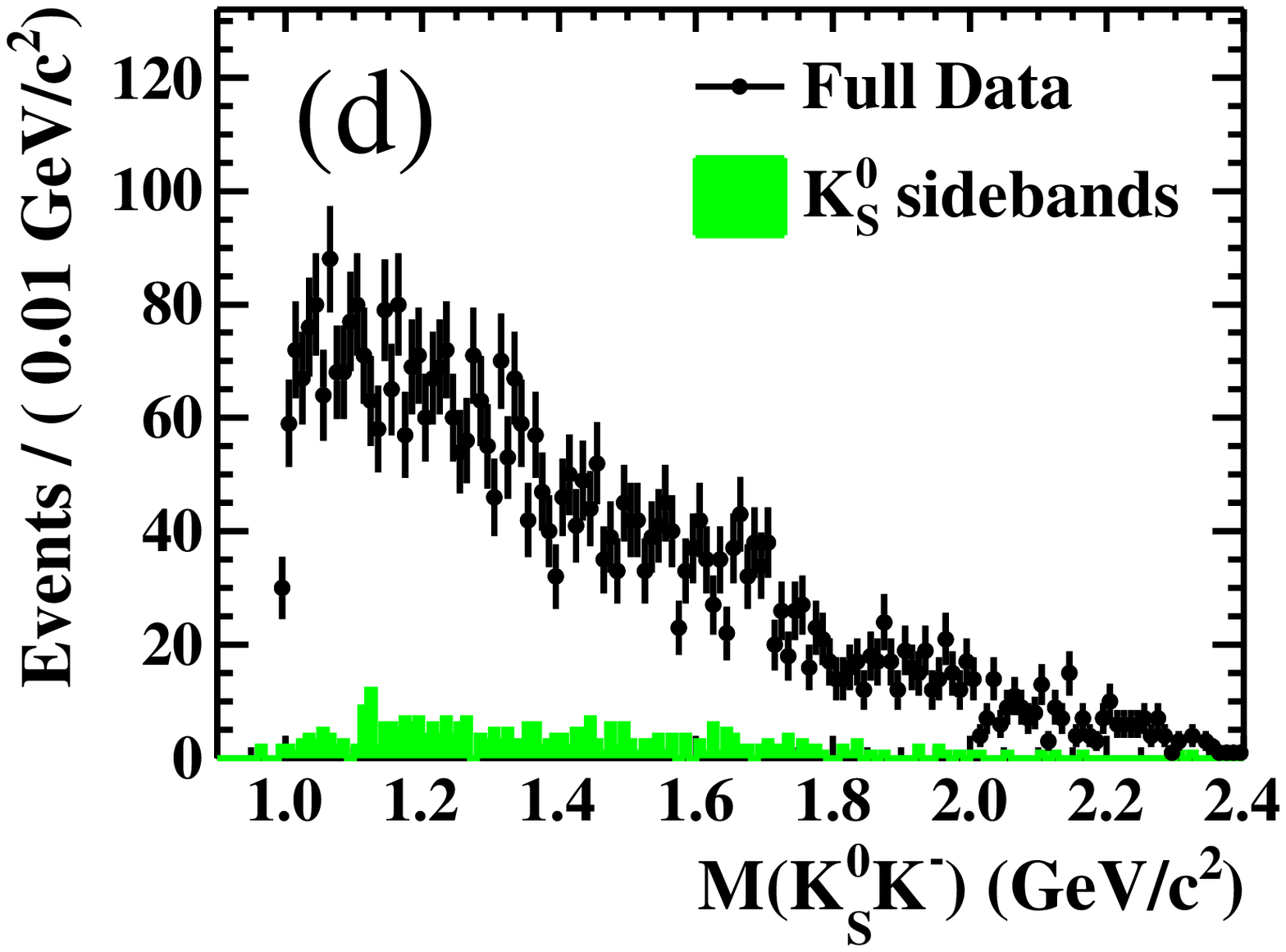}
    \end{overpic}}
    \vskip -0.2cm
    {\caption{  Invariant mass distributions of (a) $pK^{0}_{S}$, (b) $\bar{n}K^{-}$, (c) $p\bar{n}$ and (d) $K^0_SK^-$.
    The dots with error bars show the full experimental data,
    the green shaded histograms are the contributions from the normalized $\kshort$ mass sidebands events.
    }
   \label{mTheta}}
   \end{figure}

To investigate other two-body invariant mass distributions, we apply the further requirement $|M(pK^{-}/nK^{0}_{S})-1.5195|>0.025~\hbox{GeV}/c^{2}$ to veto the $\Lambda(1520)$ resonance.
The $p\kshort$, $nK^+$, $p\bar{n}$ and $\kshort K^-$ invariant mass spectra in the full data are
shown in Figs.~\ref{mTheta}~(a-d). No significant structures are visible.

\section{\boldmath Cross section measurement}

The Born cross section is calculated using:
\begin{equation}
\label{sigma-B}
\sigma_{\rm B}=\frac{N_{\rm sig}}{\mathcal{L}_{\rm int}(1+\delta)\frac{1}{|1-\Pi|^2}\epsilon\mathcal{B}},
\end{equation}
where $N_{\rm sig}$ is the number of signal events, $\mathcal{L}_{\rm int}$
is the integrated luminosity, $1+\delta$ is the
radiative correction factor obtained from a QED calculation with $1\%$
accuracy~\cite{radiator}, $\frac{1}{|1-\Pi|^2}$ is the vacuum
polarization factor~\cite{vac2,vacweb},  $\epsilon$ is the detection efficiency from the
PHSP MC simulation, $\mathcal{B}$ is the product of intermediate branching
fractions, \emph{i.e.}
$\mathcal{B}(K^{0}_{S}\rightarrow\pi^{+}\pi^{-})$ for
$e^+e^-\rightarrow pK^0_S\bar{n}K^-$,
$\mathcal{B}[\Lambda(1520)\rightarrow~pK^{-}/nK_{S}^0] \times
\mathcal{B}(K^{0}_{S}\rightarrow\pi^{+}\pi^{-})$ for
$\ee\to\Lambda(1520)\bar{n}K^{0}_{S}/\bar{p}K^+$, and
$\mathcal{B}[\Lambda(1520)\rightarrow~pK^{-}] \times
\mathcal{B}[\bar{\Lambda}(1520)\rightarrow~\bar{n}K_{S}^0] \times
\mathcal{B}(K^{0}_{S}\rightarrow\pi^{+}\pi^{-})$ for
$\ee\to\Lambda(1520)\bar{\Lambda}(1520)$.  The branching fractions
$\mathcal{B}(K^{0}_{S}\rightarrow\pi^+\pi^-)$,
$\mathcal{B}[\Lambda(1520) \to pK^{-}]$, and
$\mathcal{B}[\Lambda(1520) \to n\kshort]$ are 0.692, 0.225, and
0.1125~\cite{PDG}, respectively.  All calculated Born cross
sections or the 90\% C.L. upper limits on the Born cross sections are
summarized in Table~\ref{tab:neu} for the $pK^{0}_{s}\bar{n}K^-$ mode
and Table~\ref{tab:L1520} for the $\Lambda(1520)$ modes.

\begin{figure}[htbp]
   \centering
   \hskip -0.4cm \mbox{
   \begin{overpic}[scale=0.35,angle=0]{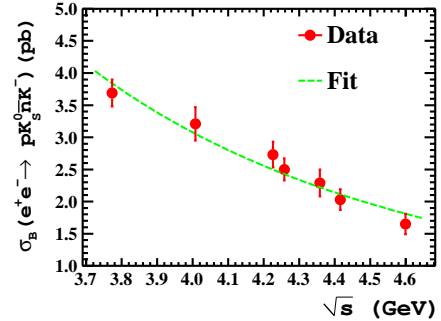}
   \put(45,45){{\LARGE }}
   \end{overpic}}
   \hskip 0.5cm
   \parbox[1cm]{8.5cm}
   {\caption{(Color online)
   Distribution of $\sigma_{\rm B}(e^+e^- \to p\kshort\bar{n}K^{-})$ versus c.m. energy.
   The dots with error bars, which show the sum in quadrature of the statistical and uncorrelated systematic uncertainties described in Section~\ref{section7}, represent data.
   The dashed line shows the fit result.
   \label{fig:section-neu}}
   }
\end{figure}

The Born cross sections of $e^+e^-\rightarrow pK^0_S\bar{n}K^-$ are
shown in Fig.~\ref{fig:section-neu} at c.m.\ energies between 3.773 and
4.6 GeV.  We fit the $1/s^{k}$ dependence of the cross sections, as
shown in Fig.~\ref{fig:section-neu} with a dashed line.  The fit gives
$k=1.9\pm0.1\pm0.2$ with the goodness of the fit $\chi^{2}/ndf=3.8/5$,
where the first uncertainty is statistical and the
second systematic.

\begin{figure}[htbp]
   \centering
   \hskip -0.4cm \mbox{
   \begin{overpic}[scale=0.35,angle=0]{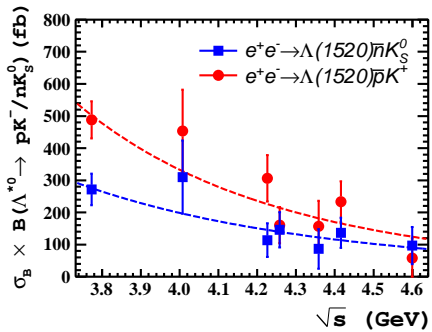}
   \put(45,45){{\LARGE }}
   \end{overpic}}
   \hskip 0.5cm
   \parbox[1cm]{8.5cm}
   {\caption{(Color online)
    Distributions of $\sigma_{\rm B}[e^+e^- \to \Lambda(1520) \bar{n}K^{0}_{S}]\times \BR[\Lambda(1520) \to p K^-]$,
    and $\sigma_{\rm B}[e^+e^- \to \Lambda(1520) \bar{p}K^{+}]\times\BR[\Lambda(1520) \to n K_S^0]$
    versus c.m. energy.
    The dots with error bars, which are the combined statistical and uncorrelated systematic uncertainties described in Section~\ref{section7}, represent data.
    The blue and red dashed lines are the fits to the cross sections
    of $e^+e^- \to \Lambda(1520) \bar{n}K^{0}_{S}$ and $e^+e^- \to
    \Lambda(1520) \bar{p}K^{+}$, respectively.}
   \label{fig:cross-section-cor}}
\end{figure}

The distributions of $\sigma[e^+e^- \to \Lambda(1520)
\bar{n}K^{0}_{S}] \times \BR[\Lambda(1520) \to p K^-]$ and
$\sigma[e^+e^- \to \Lambda(1520) \bar{p}K^{+}] \times
\BR[\Lambda(1520) \to n K_S^0]$ versus c.m.\ energy are shown in
Fig.~\ref{fig:cross-section-cor}.  A fit with the functional form $\sigma^0/s^{k}$ to each measured
Born cross section is performed, as shown in
Fig.~\ref{fig:cross-section-cor} with the dashed lines, where
$\sigma^0$ is a constant and $k$ is a free parameter.
The fits yield $k=2.8\pm0.9\pm0.2$ ($\chi^{2}/ndf=2.0/5$) and $3.5\pm0.6\pm0.2$ ($\chi^{2}/ndf=5.6/5$)
for the Born cross sections of $e^+e^- \to \Lambda(1520) \bar{n}K^{0}_{S}$
and $e^+e^- \to \Lambda(1520) \bar{p}K^{+}$, respectively,
where the first uncertainties are statistical and the second systematic.

\section{\boldmath Systematic uncertainties for Born cross sections}
\label{section7}

Sources of systematic uncertainties considered in the cross-section measurements are
the integrated luminosity measurement, tracking, PID,
$K^{0}_{S}$ reconstruction, $K^{0}_{S}$ mass window, kinematic fit for the
$\Lambda(1520)$ modes, MC generator, fitting procedure,
radiative correction, vacuum polarization, and decays of intermediate states.
The systematic uncertainties from the different sources for
all modes are summarized in Table~\ref{tab:syserror}, and the
total systematic uncertainty is obtained by adding all contributions
in quadrature assuming that each source is independent.  Detailed
descriptions of the estimates of the systematic uncertainties are
listed in the following subsections.

\begin{table*}[htbp]
  \caption{  Systematic uncertainties (\%) in the $e^+e^-\to p\kshort\bar{n}K^-$, $\Lambda(1520)\bar{n}K^{0}_{S}$,
   $\Lambda(1520)\bar{p}K^{+}$, and $\Lambda(1520)\bar{\Lambda}(1520)$ cross-section measurements.
  \label{tab:syserror}}
  \centering
  \begin{tabular*}{\textwidth}{@{\extracolsep{\fill}}l|c|ccccccc}
    \hline
    \hline
    \multicolumn{2}{l|}{$\sqrt{s}~$(GeV)}    & 3.773 & 4.008 & 4.226 & 4.258 & 4.358 & 4.416 & 4.600  \\ \hline
    \multicolumn{2}{l|}{Luminosity} & 0.5  & 1.0  & 1.0  & 1.0  & 1.0  & 1.0  & 1.0 \\  \hline
    \multicolumn{2}{l|}{Tracking}   &  3.0    &  3.0    &   3.0   &  3.0  &  3.0    &  3.0    &  3.0   \\  \hline
    \multicolumn{2}{l|}{PID [for $\Lambda(1520)$ modes]} &  3.0 & 3.0  & 3.0  & 3.0 & 3.0  & 3.0  &  3.0 \\  \hline
    \multicolumn{2}{l|}{$K^0_{S}$ reconstruction}        & 2.3  & 2.3  & 2.3  & 2.3 & 2.3  &  2.3 & 2.3  \\  \hline
    \multirow{2}*{$\kshort$ mass window} & $p\kshort\bar{n}K^-$ mode & 1.5  & 1.5  & 1.5  & 1.5 & 1.5  & 1.5  & 1.5\\
                                         & $\Lambda(1520)$ modes     & 0.8  & 0.8  & 0.8  & 0.8 & 0.8  & 0.8  & 0.8\\  \hline
    \multicolumn{2}{l|}{1C kinematic fit [for $\Lambda(1520)$ modes]}   &1.3 & 1.3 & 1.2 & 1.2 & 1.3 & 1.3 & 1.4 \\  \hline

    \multirow{2}*{MC generator}
                & $\Lambda(1520)\bar{n}\kshort$      &  3.3  &  3.3  &  3.3  & 3.3  &  3.3  &  3.3  &  3.3  \\
                & $\Lambda(1520)\bar{p}K^+$          &  1.5  &  1.5  &  1.5  & 1.5  & 1.5   & 1.5   &  1.5  \\
                & $\Lambda(1520)\bar{\Lambda}(1520)$ &  5.5  &  5.5  &  5.5  & 5.5  &  5.5  &  5.5  &  5.5  \\  \hline
    \multirow{4}*{Fit procedure} & $p\kshort\bar{n}K^-$                & 3.1 & 2.8 & 2.9 & 3.1 & 2.9 & 3.4 & 2.8  \\
                                & $\Lambda(1520)\bar{n}K^{0}_{S}$    & 2.4 & 2.0 & 2.0 & 2.0 & 2.0 & 2.0 & 2.0  \\                                & $\Lambda(1520)\bar{p}K^{+}$        & 2.5 & 3.1 & 3.1 & 3.1 & 3.1 & 3.1 & 3.1  \\
                                & $\Lambda(1520)\bar{\Lambda}(1520)$ & 4.4 & 3.9 & 3.9 & 3.9 & 3.9 & 3.9 & 3.9  \\ \hline
    \multirow{3}*{Radiative correction}  & $p\kshort\bar{n}K^-$ mode & 1.9 & 1.6 & 1.3 & 1.3 & 1.4 & 1.7 & 1.3  \\
                                         & $\Lambda(1520)$ modes   & 1.6 & 1.2 & 1.5 & 1.4 & 1.9 & 1.8 & 2.0  \\  \hline
    \multicolumn{2}{l|}{Intermediate decay [for $\Lambda(1520)$ modes]} & 2.2 & 2.2  & 2.2  & 2.2 & 2.2  & 2.2  &   2.2   \\ \hline
    \multirow{4}*{Total}        &$p\kshort\bar{n}K^-$                & 5.5  & 5.3  & 5.3  & 5.4  & 5.3  & 5.7  & 5.2 \\
                                & $\Lambda(1520)\bar{n}K^{0}_{S}$    & 7.1  & 6.9  & 7.0  & 6.9  & 7.1  & 7.0  & 7.1 \\
                                & $\Lambda(1520)\bar{p}K^{+}$        & 6.5  & 6.7  & 6.7  & 6.7  & 6.9  & 6.8  & 6.9 \\
                                & $\Lambda(1520)\bar{\Lambda}(1520)$ & 9.1  & 8.9  & 8.9  & 8.9  & 9.0  & 9.0  & 9.0\\
    \hline
    \hline
  \end{tabular*}
\end{table*}

\subsection{Integrated luminosity, tracking, and PID}
The luminosity is measured using large-angle Bhabha scattering events with a total uncertainty of
less than $1.0\%$~\cite{lum3770,lum3770new,luminosity}, which is taken as its systematic
uncertainty at each c.m.\ energy.

Using the control samples of $\ee\rightarrow
p\bar{p}\pi^{+}\pi^{-}$ at $\sqrt{s} > 4$ GeV and $J/\psi\rightarrow
K^0_{S}K^{\pm}\pi^{\mp}$ events, the tracking efficiency difference
between MC simulation and data is found to be $2.0\%$ for each proton
and $1.0\%$ for each kaon.
Not counting the two charged pions from $K^{0}_{S}$ decays,
there are two charged tracks ($p,~K$), and the
uncertainty in the tracking efficiency is $3.0\%$.

Based on the measurements of the particle identification efficiencies
of protons from $\ee\to p\bar{p}\pi^{+}\pi^{-}$ events and kaons from
$\ee\to K^{+}K^{-}\pi^{+}\pi^{-}$ events, the difference between data
and MC simulation yields uncertainties of $1.0\%$ for each proton and
$2.0\%$ for each kaon.
Thus, a total uncertainty associated with the PID of 3.0\%
is assigned for the $\Lambda(1520)$ modes.

\subsection{\boldmath $\kshort$ reconstruction and $\kshort$ mass
  window}
The $K^0_{S}$ reconstruction efficiency is studied using two control samples: $\jpsi\to
K^{*}(892)^{\pm}K^{\mp}\to K^0_{S}\pi^{\pm} K^{\mp}$ and $\jpsi\to\phi
K^{0}_{S}K^{\pm}\pi^{\mp} $.  The difference in the $\kshort$ reconstruction efficiency
between the MC simulation and the data is 1.2$\%$~\cite{Matian}.
Considering the additional PID requirements for two opposite-charged
pions from $\kshort$ decays in our analysis, the systematic
uncertainty for $\kshort$ reconstruction is conservatively taken as
$2.3\%$, where the PID efficiency difference of 1\% for each pion
between MC and data is included~\cite{2014DDstar}.

The uncertainty attributed to the $\kshort$ mass window requirement,
which originates from the mass resolution difference between the data and
the MC simulation, is estimated using $|\varepsilon_{\rm
  data}-\varepsilon_{\rm MC}|/\varepsilon_{\rm data}$, where
$\varepsilon_{\rm data}$ is the efficiency of applying the $\kshort$
mass window requirement by extracting $\kshort$ signal in the $\pi^+\pi^-$
invariant mass spectrum of the data at each c.m.\ energy, and
$\varepsilon_{\rm MC}$ is the analogous efficiency from the MC
simulation.  The difference between the data and the MC simulation is
considered as the systematic uncertainty at each c.m.\ energy.

\subsection{Kinematic fit and MC generator}

A correction is applied to the track helix parameters in the MC
simulation to make the $\chi^2$ distribution of the 1C kinematic fit from the MC simulation
agree better with data~\cite{correction}.  The
difference between the efficiencies with and without the correction is
taken as the systematic uncertainty.  In this analysis, the detection
efficiencies from the MC samples with the corrected track helix
parameters are taken as nominal results.

The detection efficiencies are obtained from the PHSP MC samples for
$\ee\to\Lambda(1520)\bar{n}\kshort$ and $\Lambda(1520)\bar{p}K^+$.
To estimate the uncertainty attributed to the MC generator, we
determine the efficiency correction factor by comparing the Dalitz
plots between the data and the MC simulation at the c.m. energy of 3.773
GeV, and the correction factor is assigned to the other
c.m. energies due to the limited statistics.  The relative differences in the
efficiency with and without correction are 3.3\% and 1.5\% for $\ee
\to \Lambda(1520)\bar{n}\kshort$ and $\ee \to
\Lambda(1520)\bar{p}K^{+}$, respectively, which are taken as the
systematic uncertainties due to the MC generator at all of c.m. energies.

For $e^+e^-\rightarrow\Lambda(1520)\bar{\Lambda}(1520)$, different
MC samples with angular distributions of $1+ \cos^2\theta$ and
$1-\cos^2\theta$ are generated, where $\theta$ is the polar angle
of $\Lambda(1520)$ in $e^+e^-$ c.m.\ frame.
The largest difference of 5.5\%, compared to the PHSP MC efficiency,
is taken as the systematic uncertainty attributed to the MC generator.

\subsection{Fit procedure}
Signal yields are determined from the fits to
the $M^{\rm rec}(p\kshort K^-)$ spectra for the $p\kshort\bar{n}K^-$ mode and the
$M(pK^{-})$ or $M(nK^{0}_{S})$ spectra for $\Lambda(1520)$ modes.
Alternative signal and background shapes as well as multiple fit ranges are
used to estimate the systematic uncertainty in the fit procedure.
We generated simulated pseudoexperiments out of the fit to the data with
alternative shapes and fitted them back using the nominal model.
Any deviation of the pull distributions from the normal gives the systematic effect.
\vspace{-1.0em}
\subsubsection{ $p\kshort\bar{n}K^-$ mode}
\vspace{-1.0em}
  \begin{itemize}
    \item[1)]
    {\it Signal shape:} In the nominal fit, the $\bar{n}$ signal shape is obtained from the MC simulation directly convolved with a Gaussian function. Alternatively, the
    incoherent sum of a Gaussian function and a Novosibirsk function~\cite{Novosibirsk}
    is taken as the $\bar{n}$ signal shape.

  \item[2)] {\it Background shape:} The background shape without
    $\kshort$ mass side-band events is described by a first-order
    polynomial function in the nominal fit, and a second-order polynomial function
    is used to estimate the systematic uncertainty due to background shape.

  \item[3)] {\it Fit range:} In the nominal fit, the fit range
    is [0.80, 1.10] GeV/$c^{2}$. The largest difference between the
    nominal fit and the fit with ranges varied to [0.805, 1.095] or
    [0.795, 1.115] GeV/$c^{2}$ is taken as the systematic
    uncertainty due to the fitting range.
  \end{itemize}

  Assuming that all of the above sources are independent,
  the systematic uncertainties associated with the fit procedure are the
  quadrature sum of above three sources.

\subsubsection{ $\Lambda(1520)$ modes}
Since only a few $\Lambda(1520)$ signal events are
observed in each $\Lambda(1520)$ mode at each c.m.\ energy in the
$XYZ$ data, we use the full $XYZ$ data to estimate the uncertainty associated with
the fit procedure for each $XYZ$ data sample.

  \begin{itemize}
  \item[1)] {\it Signal shape:} In the nominal fit, the $\Lambda(1520)$ signal is
    described by a D-wave relativistic BW function convolved with a
    Gaussian function. Alternatively, the
    $\Lambda(1520)$ signal shape is obtained from the signal MC
    simulated shape convolved with a Gaussian function.

  \item[2)] {\it Background shape:} In the nominal fit, the background
    shape is described by an ARGUS function~\cite{ref:argus}.  To
    estimate the uncertainty due to background shape, we use the
    alternative parameterized exponential function
    \begin{equation}
    \label{equbkg:L1520}
    f(M)=
    \begin{cases}
    0,                                      &\text{$M<M_0$}  \\
    (M-M_0)^{p}e^{c_1(M-M_0)+c_2(M-M_0)^2}, &\text{$M\geqslant M_0$}
    \end{cases}
    \end{equation}
    as the background shape, where $M_0$ is the threshold limit of the mass distributions, and $p$, $c_1$, and $c_2$ are free parameters.

  \item[3)] {\it Fit range:} In the nominal fit, the fit range
    is [1.41, 1.81] GeV/$c^{2}$. Changing the fit range to
    [1.41, 1.79], [1.41, 1.80], [1.41, 1.82] or [1.41, 1.83]
    GeV/$c^{2}$, the largest change of signal yields with respect to the nominal value
    is taken as the systematic uncertainty due to the fit range.
  \end{itemize}

  Assuming that all of the above sources are independent and adding
  them in quadrature, we obtain the systematic uncertainties
  associated with the fit procedure
  for $e^+e^-\rightarrow\Lambda(1520)\bar{n}K^{0}_{S}$,
  $\Lambda(1520)\bar{p}K^+$, and
  $\Lambda(1520)\bar{\Lambda}(1520)$, respectively.

\subsection{Radiative correction, vacuum polarization and intermediate decays}

The line shape of the cross section for $\ee \to p\kshort\bar{n}K^-$ affects the radiative
correction factor ($1+\delta$) and detection efficiency ($\epsilon$).
For our nominal results, the dependence of the Born cross sections on
the c.m.\ energy are $\sigma_{\rm B}\varpropto 1/s^{1.9}$, $1/s^{2.8}$ and
$1/s^{3.5}$ for $e^+e^-\rightarrow p\kshort\bar{n}K^{-}$,
$\Lambda(1520)\bar{n}K^{-}$, and
$\Lambda(1520)\bar{p}\kshort$, respectively.
The dependence is assumed to be $1/s^{3}$ for $\ee \to
\Lambda(1520)\bar{\Lambda}(1520)$
due to the limited statistics.
We change the energy dependence to $1/s$ for the above processes, and
the largest difference in $(1+\delta)\epsilon$ among the modes is
conservatively taken as the systematic uncertainty.

The vacuum polarization factor is calculated
with an uncertainty of less than 0.1$\%$~\cite{vac2}, which is negligible compared with other sources of uncertainties.

The uncertainties of $\BR(\Lambda(1520)\rightarrow pK^-/n\kshort)$
and $\BR(K^0_{S}\rightarrow\pi^{+}\pi^{-})$ are $2.2\%$ and
$0.07\%$~\cite{PDG}, respectively.  Therefore, the uncertainty from
the decays of the intermediate states is $2.2\%$ for $\Lambda(1520)$
modes, and is disregarded in the $p\kshort\bar{n}K^-$ mode.

\section{\boldmath Systematic uncertainties for fits to the cross-section distributions}
The systematic uncertainty in the measured cross section is divided into two
categories: the uncorrelated part among the different c.m.\ energies,
which comes from the fit to the $\bar{n}$ or $\Lambda(1520)$ mass
spectrum to determine the signal yields, and the correlated part,
which includes all other uncertainties common for the whole data set.
By including the uncorrelated uncertainty in the fit to the cross-section
distributions, the systematic uncertainties in the parameter $k$ are
estimated to be 8.6\%, 6.5\%, and 4.3\% for the decays of $\ee \to
p\kshort\bar{n}K^-$, $\Lambda(1520)\bar{n}\kshort$, and
$\Lambda(1520)\bar{p}K^+$, respectively.

\section{\boldmath Summary and discussion}

In summary, we study for the first time the processes $e^+e^-\rightarrow pK^{0}_{S}\bar{n}K^{-}$,
$\Lambda(1520)\bar{n}K^0_S$, $\Lambda(1520)\bar{p}K^+$, and
$\Lambda(1520)\bar{\Lambda}(1520)$ using data samples with a total
integrated luminosity of 7.4~fb$^{-1}$ collected with the BESIII
detector at c.m.\ energies of 3.773, 4.008, 4.226, 4.258, 4.358, 4.416, and
4.600 GeV.
The Born cross sections of $\ee \to p\kshort\bar{n}K^-$ are measured, and no
structure in the cross section line shape between 3.773 and 4.60 GeV is visible.

Furthermore, $\Lambda(1520)$ signals are observed in the $pK^-$ and $n
K^{0}_{S}$ invariant mass spectra for the processes $e^+e^-\rightarrow
\Lambda(1520)\bar{n}K^{0}_{S}$ and $\Lambda(1520)\bar{p}K^{+}$
with statistical significances equal to or greater than 3.0$\sigma$,
and the corresponding Born cross sections are measured.  For $e^+e^-\rightarrow\Lambda(1520)
\bar{\Lambda}(1520)$, the statistical significances are less than
3.0$\sigma$, and the 90\% C.L.\ upper limits on the Born cross sections are
determined.  No other significant structure is found in the $pK^-$,
$nK^{0}_S$, $pK^0_{S}$, $nK^+$, $p\bar{n}$ or $\kshort K^-$ invariant
mass spectra in any of the data samples.

As a consequence, no light hadron decay modes of $Y$ states or conventional
charmonium resonances are observed in our analysis.
However, we note that there is an evident difference in line shape and magnitude of
the measured cross sections between $\ee\to\Lambda(1520)\bar{n}K^{0}_{S}$ and
$\ee\to\Lambda(1520)\bar{p}K^+$
(the statistical significance of the cross-section difference
is 3.1$\sigma$ at the c.m. energy of 3.770 GeV).
Such an isospin violating effect may be due to the interference between $I=1$
and $I=0$ final states.  The final states
$\Lambda(1520)\bar{n}K^{0}_{S}$ and $\Lambda(1520)\bar{p}K^+$ can
be produced from $pK^-$ and $\bar{n}K^0_S$ systems either in $I=1$ or
$I=0$ states, namely, excited $\Sigma^*$ or $\Lambda^*$ states.  These
two states can decay into both $pK^-$ and $\bar{n}K^0_S$ final states,
but with a sign difference from Clebsch-Gordan coefficients.  Another possible
approach is $\ee \to K^*\bar{K}$ with highly excited $K^*$ decays into $\Lambda^*\bar{p}$
or $\Lambda^*\bar{n}$.  The $K^*\bar{K}$ system can be produced from $I=1$
(excited $\rho$) or $I=0$ (excited $\omega$ or $\phi$) states,
where the interference effect can occur.  If the final state is
$p\bar{N}^*$ or $n\bar{N}^*$, a similar pattern could be observed.
More experimental data are desirable to confirm these interpretations and
speculations in the future.

\section*{Acknowledgement}
The BESIII collaboration thanks the staff of BEPCII and the IHEP computing center for their strong support. This work is supported in part by National Key Basic Research Program of China under Contract No. 2015CB856700; National Natural Science Foundation of China (NSFC) under Contracts Nos. 11335008, 11425524, 11625523, 11635010, 11735014, 11705006; the Chinese Academy of Sciences (CAS) Large-Scale Scientific Facility Program; the CAS Center for Excellence in Particle Physics (CCEPP); Joint Large-Scale Scientific Facility Funds of the NSFC and CAS under Contracts Nos. U1532257, U1532258, U1732263; CAS Key Research Program of Frontier Sciences under Contracts Nos. QYZDJ-SSW-SLH003, QYZDJ-SSW-SLH040; 100 Talents Program of CAS; INPAC and Shanghai Key Laboratory for Particle Physics and Cosmology; German Research Foundation DFG under Contracts Nos. Collaborative Research Center CRC 1044, FOR 2359; Istituto Nazionale di Fisica Nucleare, Italy; Koninklijke Nederlandse Akademie van Wetenschappen (KNAW) under Contract No. 530-4CDP03; Ministry of Development of Turkey under Contract No. DPT2006K-120470; National Science and Technology fund; The Swedish Research Council; U. S. Department of Energy under Contracts Nos. DE-FG02-05ER41374, DE-SC-0010118, DE-SC-0010504, DE-SC-0012069; University of Groningen (RuG) and the Helmholtzzentrum fuer Schwerionenforschung GmbH (GSI), Darmstadt.


\begin{thebibliography}{**}
\bibitem{2008ss}
  S.~Godfrey and S.~L.~Olsen,
  Ann.\ Rev.\ Nucl.\ Part.\ Sci.\  {\bf 58}, 51 (2008).

\bibitem{2011nb}
  N.~Brambilla {\it et al.},
  Eur.\ Phys.\ J.\ C {\bf 71}, 1534 (2011).

\bibitem{2014nb}
  N.~Brambilla {\it et al.},
  Eur.\ Phys.\ J.\ C {\bf 74}, 2981 (2014).

\bibitem{2005rm}
  B.~Aubert {\it et al.} [BaBar Collaboration],
  Phys.\ Rev.\ Lett.\  {\bf 95}, 142001 (2005).

\bibitem{Lees:2012cn}
  J.~P.~Lees {\it et al.} [BaBar Collaboration],
  Phys.\ Rev.\ D {\bf 86}, 051102 (2012).

\bibitem{2006kg}
  Q.~He {\it et al.} [CLEO Collaboration],
  Phys.\ Rev.\ D {\bf 74}, 091104 (2006).

\bibitem{2007sj}
  C.~Z.~Yuan {\it et al.} [Belle Collaboration],
  Phys.\ Rev.\ Lett.\  {\bf 99}, 182004 (2007).

\bibitem{ref6}
  M.~Ablikim {\it et al.} [BESIII Collaboration],
  Phys.\ Rev.\ Lett.\  {\bf 110}, 252001 (2013).

\bibitem{ref7}
  Z.~Q.~Liu {\it et al.} [Belle Collaboration],
  Phys.\ Rev.\ Lett.\  {\bf 110}, 252002 (2013).

\bibitem{2006ss}
  X.~H.~Mo, G.~Li, C.~Z.~Yuan, K. L. He, H. M. Hu, J. H. Hu, P. Wang, and Z. Y. Wang,
  Phys.\ Lett.\ B {\bf 640}, 182 (2006).


\bibitem{ddpi-bes}
  M. Ablikim {\it et al.} (BESIII Collaboration), Evidence of a resonant structure in the $e^+e^-\to\pi^+D^0D^{*-}$ cross section between 4.05 and 4.60 GeV, arXiv:1808.02847

\bibitem{BESIII}
  M.~Ablikim {\it et al.} [BESIII Collaboration],
  Nucl.\ Instrum.\ Meth.\ A {\bf 614}, 345 (2010).

\bibitem{cmsenergy}
  M.~Ablikim {\it et al.} [BESIII Collaboration],
  Chin.\ Phys.\ C {\bf 40}, 063001 (2016).

\bibitem{lum3770}
  M.~Ablikim [BESIII Collaboration],
  Chin.\ Phys.\ C {\bf 37}, 123001 (2013).

\bibitem{lum3770new}
  M.~Ablikim {\it et al.} [BESIII Collaboration],
  Phys.\ Lett.\ B {\bf 753}, 629 (2016).

\bibitem{luminosity}
  M.~Ablikim {\it et al.} [BESIII Collaboration],
  Chin.\ Phys.\ C {\bf 39}, 093001 (2015).

\bibitem{geant4}
  S.~Agostinelli {\it et al.} [GEANT4 Collaboration],
  Nucl.\ Instrum.\ Meth.\ A {\bf 506}, 250 (2003).

\bibitem{boost}
  Z. Y. Deng {\it et al.}, Chin. Phys. C {\bf 30}, 371 (2006).

\bibitem{photos}
  P.~Golonka and Z.~Was,
  Eur.\ Phys.\ J.\ C {\bf 45}, 97 (2006).

\bibitem{kkmc:pingrg}
  R.~G.~Ping,
  Chin.\ Phys.\ C {\bf 38}, 083001 (2014).

\bibitem{kkmc1}
  S.~Jadach, B.~F.~L.~Ward, and Z.~Was,
  Comput.\ Phys.\ Commun.\  {\bf 130}, 260 (2000).

\bibitem{kkmc2}
  S.~Jadach, B.~F.~L.~Ward, and Z.~Was,
  Phys.\ Rev.\ D {\bf 63}, 113009 (2001).

\bibitem{besevtgen}
  R.~G.~Ping,
  Chin.\ Phys.\ C {\bf 32}, 599 (2008).

\bibitem{evtgen}
  D.~J.~Lange,
  Nucl.\ Instrum.\ Meth.\ A {\bf 462}, 152 (2001).

\bibitem{PDG}
M. Tanabashi {\it et al.} [Particle Data Group], Phys. Rev. D {\bf 98}, 030001 (2018).

\bibitem{Chen:2000tv}
  J.~C.~Chen, G.~S.~Huang, X.~R.~Qi, D. H. Zhang, and Y. S. Zhu,
  Phys.\ Rev.\ D {\bf 62}, 034003 (2000).

\bibitem{pythia}
  T.~Sj{\"{o}}strand {\it et al.},
  Comput.\ Phys.\ Commun.\  {\bf 191}, 159 (2015).

\bibitem{sec-vtx}
  M.~Xu {\it et al.},
  Chin.\ Phys.\ C {\bf 33}, 428 (2009).

\bibitem{ref:argus}
  H.~Albrecht {\it et al.} [ARGUS Collaboration],
  Phys.\ Lett.\ B {\bf 340}, 217 (1994).

\bibitem{width-dependent}
  S.~Kopp {\it et al.} [CLEO Collaboration],
  Phys.\ Rev.\ D {\bf 63}, 092001 (2001).

\bibitem{radiator}
  E.~A.~Kuraev and V.~S.~Fadin,
  Sov.\ J.\ Nucl.\ Phys.\  {\bf 41}, 466 (1985).

\bibitem{vac2}
  S.~Actis {\it et al.},
  Eur.\ Phys.\ J.\ C {\bf 66}, 585 (2010).

\bibitem{vacweb}
  F. Jegerlehner, [Online; accessed March-2018],
  \url{http://www-com.physik.hu-berlin.de/~fjeger/}.


\bibitem{Matian}
  M.~Ablikim {\it et al.} [BESIII Collaboration],
  Phys.\ Rev.\ D {\bf 91}, 112008 (2015).

\bibitem{2014DDstar}
  M.~Ablikim {\it et al.} [BESIII Collaboration],
  Phys.\ Rev.\ Lett.\  {\bf 112}, 022001 (2014).

\bibitem{correction}
  M.~Ablikim {\it et al.} [BESIII Collaboration],
  Phys.\ Rev.\ D {\bf 87}, 012002 (2013).

\bibitem{Novosibirsk}
  H.~S.~Ahn {\it et al.},
  Nucl.\ Instrum.\ Meth.\ A {\bf 410}, 179 (1998).

\end{thebibliography}

\end{document}